\documentstyle[preprint,psfig,aps,eqsecnum,floats]{revtex}
 
\input epsf
 
\newcommand{\beq}{\begin{equation}}
\newcommand{\eeq}{\end{equation}}
\newcommand{\beqa}{\begin{eqnarray}}
\newcommand{\eeqa}{\end{eqnarray}}

\newcommand{\apnn}{\mbox{$a_{\pi\nu\bar\nu}$}}
\newcommand{\apks}{\mbox{$a_{\psi K_S}$}}

\def\epsK{\varepsilon_K}
\def\ra{\rightarrow}
\def\epspK{\varepsilon^\prime_K}
\def\ra{\rightarrow}
\def\CP{{\rm CP}}
\def\O{{\cal O}}
\def\L{{\cal L}}
\def\Re{{\cal R}e}
\def\Im{{\cal I}m}
\def\gsim{\lower.8ex\hbox{$\sim$}\kern-.75em\raise.45ex\hbox{$>$}\;}
\def\lsim{\lower.8ex\hbox{$\sim$}\kern-.8em\raise.45ex\hbox{$<$}\;}

\def\tgl{{\tilde g}}

\def\gev{{\rm GeV}}

\def\H{{\cal H}}

\def\dKM{\delta_{\rm KM}}

\def\apks{a_{\psi K_S}}
\def\apnn{a_{\pi\nu\bar\nu}}
\def\epsK{\varepsilon_K}

\def\vev#1{\langle #1 \rangle} 
\def\tgl{{\tilde g}}
 
\begin{document}
 
\draft
 
{\tighten
 
\preprint{
\vbox{
      \hbox{WIS-99/15/Mar-DPP}
      \hbox{hep-ph/9904271} }}

\renewcommand{\thefootnote}{\fnsymbol{footnote}}
 
\title{CP Violation in $B$ Decays}
\author{Yosef Nir}
\address{ \vbox{\vskip 0.truecm}
Department of Particle Physics, \\
Weizmann Institute of Science, Rehovot 76100, Israel \\
}
 
\maketitle
\thispagestyle{empty}
\setcounter{page}{0}
\begin{abstract}
In the near future, we will have the first significant experimental  
measurements of CP violation in $B$ decays. These measurements will easily 
test crucial questions such as whether the Standard Model Kobayashi-Maskawa 
phase plays a dominant role in CP violation or whether CP is an approximate 
symmetry in nature. We explain the different types of CP violation in 
$B$ decays, and the usefulness of measuring them.
We use the same formalism to describe the $\epsK$ and $\epspK$ parameters
of the neutral $K$ system and to explain the terms direct and indirect
CP violation.
We present the Standard Model predictions for the various
asymmetries. We argue that certain CP asymmetries in $B$ decays
are subject to a very clean theoretical interpretation in
terms of fundamental Lagrangian parameters. Within the Standard
Model, these asymmetries will provide very accurate measurements
of the CKM parameters. In case that deviations from the Standard
Model predictions will be found, there is enough information to
understand the nature of New Physics that is required to explain
them. We demonstrate this statement by analyzing the impact
of various Supersymmetric flavor models on CP violation.
\end{abstract}

\bigskip
\centerline{\it Based on lectures given at the school on}
\centerline{\bf Flavour and Gauge Hierarchies}
\centerline{\it NATO Advanced Study Institute, Carg\`ese, Corsica}
\centerline{\it 20 July - 1 August 1998}
\centerline{\it and at}
\centerline{\bf The First Particle Physics Winter School}
\centerline{\it KIAS, Seoul, Korea}
\centerline{\it February 22 - 26 1999} 

} 
 
\newpage
 
\section{A MODEL INDEPENDENT DISCUSSION}
\subsection{Introduction}

CP violation arises naturally in the three generation Standard Model.
The CP violation that has been measured in neutral $K$-meson decays
($\epsK$ and $\epspK$) is accommodated in the Standard Model in simple 
way \cite{KoMa}. Yet, CP violation is one of the least tested aspects 
of the Standard Model. The value of the $\epsK$ parameter \cite{CCFT}\
as well as bounds
on other CP violating parameters (most noticeably, the electric
dipole moments of the neutron, $d_N$, and of the electron, $d_e$)
can be accounted for in models where CP violation has features
that are very different from the Standard Model ones.

It is unlikely that the Standard Model provides the complete description
of CP violation in nature. First, it is quite clear that there exists
New Physics beyond the Standard Model. Almost any extension of the Standard
Model has additional sources of CP violating effects. In addition
there is a great puzzle in cosmology that relates to CP violation,
and that is the baryon asymmetry of the universe \cite{Sakh}.
Theories that explain the observed asymmetry must include new sources of 
CP violation \cite{CKN}: the Standard Model cannot
generate a large enough matter-antimatter imbalance to produce the
baryon number to entropy ratio observed in the universe today
\cite{FaSh,Gave,HuSa}.

In the near future, significant new information on CP violation
will be provided by various experiments. The main source of information
will be measurements of CP violation in various $B$ decays, particularly
neutral $B$ decays into final CP eigenstates \cite{BCP,BiSa,DuRo}.
Another piece of valuable information might come from a measurement of the
$K_L\ra\pi^0\nu\bar\nu$ decay \cite{Litt,BuBuB,BuBun,GrNi}.
For the first time, the pattern of CP violation that is predicted by the 
Standard Model will be tested. Basic questions such as whether CP is an 
approximate symmetry in nature will be answered.

It could be that the scale where new CP violating sources appear
is too high above the Standard Model scale ({\it e.g.} the GUT scale)
to give any observable deviations from the Standard Model predictions.
In such a case, the outcome of the experiments will be a (frustratingly)
successful test of the Standard Model and a significant improvement in
our knowledge of the CKM matrix.

A much more interesting situation will arise
if the new sources of CP violation appear at a scale
that is not too high above the electroweak scale. Then they might be
discovered in the forthcoming experiments. Once enough independent
observations of CP violating effects are made, we will find that there
is no single choice of CKM parameters that is consistent with
all measurements. There may even be enough information in the pattern
of the inconsistencies to tell us something about the nature of the
new physics contributions \cite{NiSia,DLN,NiQuR,GrLoR}.

The aim of this review is to explain the theoretical
tools with which we will analyze new information about CP violation.
In this chapter, we give a brief, model-independent discussion
of CP violating observables. In the next chapter,
we discuss CP violation in the Standard Model.
In the third chapter we briefly explain why CP violation is a
powerful probe of new physics. In the last chapter, we describe CP violation 
in Supersymmetric models. This discussion enables
us to elucidate the uniqueness of the Standard Model description
of CP violation and how little it has been tested so far. It further
demonstrates how the information from CP violation can help us probe
in detail models of New Physics.

\subsection{Neutral Meson Mixing}
Much of the exciting CP violation in meson decays is related to
neutral meson mixing. Before we focus on CP violation, we briefly
discuss then the physics and formalism of neutral meson mixing.
We refer specifically to the neutral $B$ meson system, but
most of our discussion applies equally well to the neutral $K$,
$B_s$ and $D$ meson systems.

Our phase convention for the CP transformation law of
the neutral $B$ mesons is defined by
\beq\label{phacon}
\CP|{B^0}\rangle=\omega_B|{\bar B^0}\rangle,\ \ \
\CP|{\bar B^0}\rangle=\omega_B^*|{B^0}\rangle,\ \ \ (|\omega_B|=1).
\eeq
Physical observables do not depend on the phase factor $\omega_B$.
An arbitrary linear combination of the neutral $B$-meson
flavor eigenstates,
\beq\label{defab}
a|{B^0}\rangle+b|{\bar B^0}\rangle,
\eeq
is governed by a time-dependent Schr\"odinger equation,
\beq\label{Schro}
i{d\over dt}\pmatrix{a\cr b\cr}=H\pmatrix{a\cr b\cr}
\equiv\left(M-{i\over2}\Gamma\right)\pmatrix{a\cr b\cr},
\eeq
for which $M$ and $\Gamma$ are $2\times2$ Hermitian matrices.

The off-diagonal terms in these matrices, $M_{12}$ and $\Gamma_{12}$,
are particularly important in the discussion of mixing and CP violation.
$M_{12}$ is the dispersive part of the transition amplitude from
$B^0$ to $\bar B^0$. In the Standard Model it arises only at
order $g^4$. In the language of quark diagrams, the leading
contribution is from box diagrams. At sufficiently high loop momentum,
$k\gg\Lambda_{\rm QCD}$, these diagrams are a very good approximation
to the Standard Model contribution to $M_{12}$. This, or any other
contribution from heavy intermediate states from new physics, is
the {\it short distance} contribution. For small loop momenta,
$k\lsim 1$ GeV, we do not expect quark hadron duality to hold.
The box diagram is a poor approximation to the contribution from
light intermediate states, namely to {\it long distance} contributions.
Fortunately, in the $B$ and $B_s$ systems, the long distance contributions
are expected to be negligible. (This is not the case for $K$ and $D$
mesons. Consequently, it is difficult to extract useful information
from the measurement of $\Delta m_K$ and from the bound on $\Delta m_D$.)
$\Gamma_{12}$ is the absorptive part of the transition amplitude.
Since the cut of a diagram always involves on-shell particles and
thus long distance physics, the cut of the quark box diagram is a poor
approximation to $\Gamma_{12}$. However, it does correctly give
the suppression from small electroweak parameters such as the weak
coupling. In other words, though the hadronic uncertainties are large
and could change the result by order $50\%$, the cut in the
box diagram is expected to give a reasonable order of magnitude estimate
of $\Gamma_{12}$. (For $\Gamma_{12}(B_s)$ it has been shown that local
quark-hadron duality holds exactly in the simultaneous limit of small
velocity and large number of colors. We thus expect an uncertainty of
$\O(1/N_C)\sim30\%$
\cite{Alek,Bene}. For $\Gamma_{12}(B_d)$ the small velocity limit
is not as good an approximation but an uncertainty of order 50\% still
seems a reasonable estimate.) New physics is not expected to affect
$\Gamma_{12}$ significantly because it usually takes place at a high
energy scale and is relevant to the short distance part only.

The light $B_L$ and heavy $B_H$ mass eigenstates are given by
\beq\label{defqp}
|{B_{L,H}}\rangle=p|{B^0}\rangle\pm q|{\bar B^0}\rangle.
\eeq
The complex coefficients $q$ and $p$ obey the normalization condition
$|q|^2+|p|^2=1$. Note that $\arg(q/p^*)$ is just an overall common
phase for $|B_L\rangle$ and $|B_H\rangle$ and has no physical significance.
The mass difference and the width difference between the
physical states are given by
\beq\label{DelmG}
\Delta m\equiv M_H-M_L,\ \ \ \Delta\Gamma\equiv\Gamma_H-\Gamma_L.
\eeq
Solving the eigenvalue equation gives
\beq\label{eveq}
(\Delta m)^2-{1\over4}(\Delta\Gamma)^2=
(4|M_{12}|^2-|\Gamma_{12}|^2),\ \ \ \ \ 
\Delta m\Delta\Gamma=4\Re(M_{12}\Gamma_{12}^*),
\eeq
\beq\label{solveqp}
{q\over p}=-{2M_{12}^*-i\Gamma_{12}^*\over
\Delta m-{i\over2}\Delta\Gamma}=-{\Delta m-{i\over2}\Delta\Gamma
\over 2M_{12}-i\Gamma_{12}}.
\eeq
In the $B$ system, $|\Gamma_{12}|\ll|M_{12}|$ (see discussion below),
and then, to leading order in $|\Gamma_{12}/M_{12}|$, (\ref{eveq}) 
and (\ref{solveqp}) can be written as
\beq\label{eveqB}
\Delta m_B=2|M_{12}|,\ \ \
\Delta\Gamma_B=\ 2\Re(M_{12}\Gamma_{12}^*)/|M_{12}|,
\eeq
\beq\label{solveqpB}
{q\over p}=-{M_{12}^*\over|M_{12}|}.
\eeq

\subsection{CP Violation in Neutral Meson Mixing}
To discuss CP violation in mixing (see below),
it is useful to write (\ref{solveqpB}) to first order in 
$|\Gamma_{12}/M_{12}|$:
\beq\label{solveqpC}
{q\over p}=-{M_{12}^*\over|M_{12}|}\left[1-{1\over2}
\Im\left({\Gamma_{12}\over M_{12}}\right)\right].
\eeq

To discuss CP violation in decay (see below), we need to consider
decay amplitudes. The CP transformation law for a final state $f$ is
\beq\label{phaconf}
\CP|{f}\rangle=\omega_f|{\bar f}\rangle,\ \ \
\CP|{\bar f}\rangle=\omega_f^*|{f}\rangle,\ \ \ (|\omega_f|)=1.
\eeq
For a final CP eigenstate $f=\bar f=f_{\CP}$, the phase factor
$\omega_f$ is replaced by $\eta_{f_{\CP}}=\pm1$, the CP eigenvalue
of the final state.
We define the decay amplitudes $A_f$ and $\bar A_f$ according to
\beq\label{defAf}
A_f=\vev{f|{\cal H}_d|B^0},\ \ \ \bar A_f=\vev{f|{\cal H}_d|\bar B^0},
\eeq
where ${\cal H}_d$ is the decay Hamiltonian.

To discuss CP violation in the interference of decays with and
without mixing (see below), we introduce a complex quantity
$\lambda_f$ defined by
\beq\label{deflam}
\lambda_f\ =\ {q\over p}\ {\bar A_f\over A_f}.
\eeq

We further define the CP transformation law for the quark fields
in the Hamiltonian (a careful treatment of CP conventions
can be found in \cite{BLS}):
\beq\label{CPofq}
q\ \rightarrow\ \omega_q\bar q,\ \ \ 
\bar q\ \rightarrow\ \omega_q^*q,\ \ \ (|\omega_q|=1).
\eeq
The effective Hamiltonian that is relevant to $M_{12}$ is of the form
\beq\label{Hbtwo}
H^{\Delta b=2}_{\rm eff}\propto
e^{+2i\phi_B}\left[\bar d\gamma^\mu(1-\gamma_5)b\right]^2
+e^{-2i\phi_B}\left[\bar b\gamma^\mu(1-\gamma_5)d\right]^2,
\eeq
where $2\phi_B$ is a CP violating (weak) phase. (We use the
Standard Model $V-A$ amplitude, but the results can be generalized
to any Dirac structure.) For the $B$ system, where $|\Gamma_{12}|\ll
|M_{12}|$, this leads to
\beq\label{qpforB}
q/p=\omega_B\omega_b^*\omega_d e^{-2i\phi_B}.
\eeq
(We implicitly assumed that the vacuum insertion approximation
gives the correct sign for $M_{12}$. In general, there is a
sign($B_B$) factor on the right hand side of (\ref{qpforB}) \cite{GKN}.)
To understand the phase structure of decay amplitudes, we take as
an example the $b\rightarrow q\bar qd$ decay ($q=u$ or $c$). 
The decay Hamiltonian is of the form
\beq\label{Hdecay}
H_d\propto e^{+i\phi_f}\left[\bar q\gamma^\mu(1-\gamma_5)d\right]
\left[\bar b\gamma_\mu(1-\gamma_5)q\right]
+e^{-i\phi_f}\left[\bar q\gamma^\mu(1-\gamma_5)b\right]
\left[\bar d\gamma_\mu(1-\gamma_5)q\right],
\eeq
where $\phi_f$ is the appropriate weak phase. (Again, for simplicity
we use a $V-A$ structure, but the results hold for any Dirac structure.)
Then
\beq\label{AbarA}
\bar A_{\bar f}/A_f=\omega_f\omega_B^*\omega_b\omega_d^* e^{-2i\phi_f}.
\eeq
Eqs. (\ref{qpforB}) and (\ref{AbarA}) together imply that for a final 
CP eigenstate,
\beq\label{lamfCP}
\lambda_{f_{\CP}}=\eta_{f_{\CP}}e^{-2i(\phi_B+\phi_f)}.
\eeq

\subsection{The Three Types of CP Violation in Meson Decays}
There are three different types of CP violation in meson decays:
\begin{itemize}
\item[(i)] CP violation in mixing, which occurs when the two neutral mass
eigenstate admixtures cannot be chosen to be CP-eigenstates;
\item[(ii)] CP violation in decay, which occurs in both charged and neutral
decays, when the amplitude for a decay and its CP-conjugate process
have different magnitudes;
\item[(iii)] CP violation in the interference of decays with and
without mixing, which occurs in decays into final states that are common to
$B^0$ and $\bar B^0$. (It often occurs in combination with the other two
types but there are cases when, to an excellent
approximation, it is the only effect.)
\end{itemize}
(In cascade decays \cite{Azim,AzDu,KaSt,KaCA}, there appears a fourth 
type of CP violation
\cite{MeSi,ASS}. We do not discuss this type of CP violation here.)

{\bf (i) CP violation in mixing:}
\beq\label{inmixin}
|q/p|\neq1.
\eeq
This results from the mass eigenstates being different from the
CP eigenstates, and requires a relative phase between $M_{12}$
and $\Gamma_{12}$. For the neutral $B$ system, this effect could
be observed through the asymmetries in semileptonic decays:
\beq\label{mixexa}
a_{\rm SL}={\Gamma(\bar B^0_{\rm phys}(t)\ra\ell^+\nu X)-
\Gamma(B^0_{\rm phys}(t)\ra\ell^-\nu X)\over
\Gamma(\bar B^0_{\rm phys}(t)\ra\ell^+\nu X)+
\Gamma(B^0_{\rm phys}(t)\ra\ell^-\nu X)}.
\eeq
In terms of $q$ and $p$,
\beq\label{mixter}
a_{\rm SL}={1-|q/p|^4\over1+|q/p|^4}.
\eeq
CP violation in mixing has been observed in the neutral $K$ system
($\Re\ \epsK\neq0$).

In the neutral $B$ system, the effect is
expected to be small, $\lsim\O(10^{-2})$. The reason is that,
model independently, the effect cannot be larger than
$\O(\Delta\Gamma_B/\Delta m_B)$. The difference in width is produced
by decay channels common to $B^0$ and $\bar B^0$. The branching
ratios for such channels are at or below the level of $10^{-3}$.
Since various channels contribute with differing signs, one expects
that their sum does not exceed the individual level. Hence, we can safely
assume that $\Delta\Gamma_B/\Gamma_B=\O(10^{-2})$. On the other hand, it is
experimentaly known that $\Delta m_B/\Gamma_B\approx0.7$.

To calculate the deviation of $|q/p|$ from a pure phase (see (\ref{solveqpC})),
\beq\label{Absqp}
1-\left|{q\over p}\right|={1\over2}\Im{\Gamma_{12}\over M_{12}},
\eeq
one needs to calculate $M_{12}$ and $\Gamma_{12}$. This involves
large hadronic uncertainties, in particular in the hadronization
models for $\Gamma_{12}$.

{\bf (ii) CP violation in decay:}
\beq\label{indecay}
|\bar A_{\bar f}/A_f|\neq1.
\eeq
This appears as a result of interference among various terms in the
decay amplitude, and will not occur unless at least two terms have
different weak phases and different strong phases. CP asymmetries in
charged $B$ decays,
\beq\label{decexa}
a_{f}={\Gamma(B^+\ra f^+)-\Gamma(B^-\ra f^-)\over
\Gamma(B^+\ra f^+)+\Gamma(B^-\ra f^-)},
\eeq
are purely an effect of CP violation in decay.
In terms of the decay amplitudes,
\beq\label{decter}
a_{f^\pm}={1-|\bar A_{f^-}/A_{f^+}|^2\over1+|\bar A_{f^-}/A_{f^+}|^2}.
\eeq
CP violation in decay has been observed in the neutral $K$ system 
($\Re\ \epsK^\prime\neq0$ \cite{NAto,Esto,KTeVep}).

There are two types of phases that may appear in $A_f$ and
$\bar A_{\bar f}$. Complex parameters in any Lagrangian term that
contributes to the
amplitude will appear in complex conjugate form in the CP-conjugate
amplitude. Thus their phases appear in $A_f$ and $\bar A_{\bar f}$ with
opposite signs. In the Standard Model these phases occur only in the CKM
matrix which is part of the electroweak sector of the theory, hence these
are often called ``weak phases''. The weak phase of any single term is
convention dependent. However the difference between the weak phases in
two different terms in $A_f$ is convention independent because the phase
rotations of the initial and final states are the same for every term.
A second type of phase can appear in scattering or decay amplitudes even
when the Lagrangian is real. Such phases do not violate CP, since they
appear in $A_f$ and $\bar A_{\bar f}$ with the same sign. Their origin is
the possible contribution from intermediate on-shell states in the
decay process, that is an absorptive part of an amplitude that has
contributions from coupled channels. Usually the dominant
rescattering is due to strong interactions and hence the designation
``strong phases'' for the phase shifts so induced. Again only the
relative strong phases of different terms in a scattering amplitude
have physical content, an overall phase rotation of the entire
amplitude has no physical consequences.

Thus it is useful to write each contribution to $A$ in three parts:
its magnitude $A_i$; its weak phase term $e^{i\phi_i}$; and its strong
phase term $e^{i\delta_i}$. Then, if several amplitudes contribute to
$B\ra f$, we have
\beq\label{defAtoA}
\left|{\bar A_{\bar f}\over A_f}\right|=\left|{\sum_i A_i 
e^{i(\delta_i-\phi_i)}\over  \sum_i A_i e^{i(\delta_i+\phi_i)}}\right|.
\eeq
The magnitude and strong phase of any amplitude involve long distance
strong interaction physics, and our ability to calculate these from
first principles is limited. Thus quantities that depend only on the weak
phases are much cleaner than those that require knowledge of the
relative magnitudes or strong phases of various amplitude contributions,
such as CP violation in decay.
There is however a large literature and considerable theoretical effort
that goes into the calculation of amplitudes and strong phases . In many
cases we can only relate experiment to Standard Model parameters through
such calculations. The techniques that are used are expected to be more
accurate for $B$ decays than for $K$ decays, because of the larger $B$
mass, but theoretical uncertainty remains significant. The
calculations generally contain two parts. First the operator product
expansion and QCD perturbation theory are used to write any underlying
quark process as a sum of local quark operators with well-determined
coefficients. Then the matrix elements of the operators between the
initial and final hadron states must be calculated. This is where
theory is weakest and the results are most model dependent. Ideally
lattice calculations should be able to provide accurate determinations
for the matrix elements, and in certain cases this is already true, but
much remains to be done.

{\bf (iii) CP violation in the interference between decays
 with and without mixing:}
\beq\label{ininter}
|\lambda_{f_{\CP}}|=1,\ \ \Im\ \lambda_{f_{\CP}}\neq0.
\eeq
Any $\lambda_{f_{\CP}}\neq\pm1$ is a manifestation of CP violation.
The special case (\ref{ininter}) isolates the effects of interest since both
CP violation in decay (\ref{indecay}) and in mixing (\ref{inmixin}) lead to
$|\lambda_{f_{\CP}}|\neq1$. For the neutral $B$ system, this effect can
be observed by comparing decays into final CP eigenstates of a
time-evolving neutral $B$ state that begins at time zero as $B^0$
to those of the state that begins as $\bar B^0$:
\beq\label{intexa}
a_{f_{\CP}}={\Gamma(\bar B^0_{\rm phys}(t)\ra f_{\CP})-
\Gamma(B^0_{\rm phys}(t)\ra f_{\CP})\over
\Gamma(\bar B^0_{\rm phys}(t)\ra f_{\CP})+
\Gamma(B^0_{\rm phys}(t)\ra f_{\CP})}.
\eeq
This time dependent asymmetry is given (for $|\lambda_{f_{\CP}}|=1$) by
\beq\label{intters}
a_{f_{\CP}}=-\Im\lambda_{f_{\CP}}\sin(\Delta m_B t).
\eeq

CP violation in the interference of decays with and without
mixing has been observed for the neutral $K$ system
($\Im\ \epsK\neq0$). It is expected to be an effect of $\O(1)$
in various $B$ decays.  For such cases, the contribution from CP violation
in mixing is clearly negligible. For decays that are dominated
by a single CP violating phase (for example, $B\ra\psi K_S$ and
$K_L\ra\pi^0\nu\bar\nu$), so that the contribution from CP violation in
decay is also negligible, $a_{f_{\rm CP}}$ is cleanly interpreted
in terms of purely electroweak parameters. Explicitly,
$\Im\lambda_{f_{\CP}}$ gives the difference between the phase of
the $B-\bar B$ mixing amplitude ($2\phi_B$) and twice the phase of the
relevant decay amplitude ($2\phi_f$) (see  eq. (\ref{lamfCP})):
\beq\label{intCKM}
\Im\lambda_{f_{\CP}}=-\eta_{f_{\CP}}\sin[2(\phi_B+\phi_f)].
\eeq

\subsection{Indirect vs. Direct CP Violation}
The terms indirect CP violation and direct CP violation
are commonly used in the literature. While various authors use these
terms with different meanings, the most useful definition is the following:
\begin{itemize}
\item[(i)] {\bf Indirect CP violation} refers to CP violation in
meson decays where the CP violating phases can all be chosen to
appear in $\Delta F=2$ (mixing) amplitudes.
\item[(ii)] {\bf Direct CP violation} refers to CP violation in
meson decays where some CP violating phases necessarily
appear in $\Delta F=1$ (decay) amplitudes.
\end{itemize}

Examining eqs. (\ref{inmixin}) and (\ref{solveqp}), we learn that
CP violation in mixing is a manifestation of indirect CP violation.
Examining eqs. (\ref{indecay}) and (\ref{defAf}), we learn that
CP violation in decay is a manifestation of direct CP violation.
Examining eqs. (\ref{ininter}) and (\ref{deflam}), we learn that
the situation concerning CP violation in the interference of
decays with and without mixing is more subtle. For any single
measurement of $\Im\lambda_f\neq0$, the relevant CP violating phase
can be chosen by convention to reside in the $\Delta F=2$ amplitude
($\phi_f=0$, $\phi_B\neq0$ in the notation of eq. (\ref{lamfCP})),
and then we would call it indirect CP violation. Consider, however,
the CP asymmetries for two different final CP eigenstates (for the
same decaying meson), $f_a$ and $f_b$. Then, a non-zero difference
between $\Im\lambda_{f_a}$ and $\Im\lambda_{f_b}$ requires
that there exists CP violation in $\Delta F=1$ processes ($\phi_{f_a}-
\phi_{f_b}\neq0$), namely direct CP violation.

Experimentally, both direct and indirect CP violation have been established.
Below we will see that $\epsK$ signifies indirect CP violation
while $\epspK$ signifies direct CP violation. 

Theoretically, most models of CP violation (including the Standard Model)
have predicted that both types of CP violation exist. There is, however,
one class of models, that is {\it superweak models}
\cite{WolSW,WinSW,SWSW,WWSW}, that predict
only indirect CP violation. The measurement of $\epspK\neq0$ has
excluded this class of models.

\subsection{The $\epsK$ and $\epspK$ Parameters}
Historically, a different language from the one used by us has been
employed to describe CP violation in $K\ra\pi\pi$ and $K\ra\pi\ell\nu$
decays. In this section we `translate' the language of $\varepsilon_K$
and $\varepsilon_K^\prime$ to our notations. Doing so will make it
easy to understand which type of CP violation is related to each quantity.

The two CP violating quantities measured in neutral $K$ decays are
\beq\label{defetaij}
\eta_{00}={\vev{\pi^0\pi^0|\H|K_L}\over\vev{\pi^0\pi^0|\H|K_S}},\ \ \ 
\eta_{+-}={\vev{\pi^+\pi^-|\H|K_L}\over\vev{\pi^+\pi^-|\H|K_S}}.
\eeq
Define for $(ij)=(00)$ or $(+-)$
\beq\label{epsamp}
A_{ij}=\vev{\pi^i\pi^j|\H|K^0},\ \ \ 
\bar A_{ij}=\vev{\pi^i\pi^j|\H|\bar K^0},
\eeq
\beq\label{epslam}
\lambda_{ij}=\left({q\over p}\right)_K{\bar A_{ij}\over A_{ij}}.
\eeq
Then
\beq\label{etapqA}
\eta_{00}={1-\lambda_{00}\over1+\lambda_{00}},\ \ \ 
\eta_{+-}={1-\lambda_{+-}\over1+\lambda_{+-}}.
\eeq
The $\eta_{00}$ and $\eta_{+-}$ parameters get contributions from 
CP violation in mixing ($|(q/p)|_K\neq1$) and from the interference 
of decays with and without mixing ($\Im\lambda_{ij}\neq0$) 
at $\O(10^{-3})$ and from CP violation in decay 
($|\bar A_{ij}/A_{ij}|\neq1$) at $\O(10^{-6})$.

There are two isospin channels in $K\ra\pi\pi$ leading to final
$(2\pi)_{I=0}$ and $(2\pi)_{I=2}$ states:
\beq\label{twoisZ}
\langle\pi^0\pi^0|=\sqrt{1\over3}\langle(\pi\pi)_{I=0}|-
\sqrt{2\over3}\langle(\pi\pi)_{I=2}|,
\eeq
\beq\label{twoisT}
\langle\pi^+\pi^-|=\sqrt{2\over3}\langle(\pi\pi)_{I=0}|+
\sqrt{1\over3}\langle(\pi\pi)_{I=2}|.
\eeq
The fact that there are two strong phases allows for CP violation
in decay. The possible effects are, however, small (on top of the 
smallness of the relevant CP violating phases) because the final
$I=0$ state is dominant (this is the $\Delta I=1/2$ rule). Defining
\beq\label{defAI}
A_I=\vev{(\pi\pi)_I|\H|K^0},\ \ \ \bar A_I=\vev{(\pi\pi)_I|\H|\bar K^0},
\eeq
we have, experimentally,
\beq\label{AtwoAzer}
|A_2/A_0|\approx1/20.
\eeq
Instead of $\eta_{00}$ and $\eta_{+-}$ we may define two combinations,
$\epsK$ and $\epspK$, in such a way that the possible effects of
CP violation in decay (mixing) are isolated into $\epspK$ ($\epsK$).

The experimental definition of the $\epsK$ parameter is 
\beq\label{defepsex}
\epsK\equiv{1\over3}(\eta_{00}+2\eta_{+-}).
\eeq
To zeroth order in $A_2/A_0$, we have $\eta_{00}=\eta_{+-}=\epsK$.
However, the specific combination (\ref{defepsex}) is chosen in such a 
way that  the following relation holds to {\it first} order in $A_2/A_0$:
\beq\label{defepsth}
\epsK={1-\lambda_0\over1+\lambda_0},
\eeq
where
\beq\label{deflamz}
\lambda_0=\left({q\over p}\right)_K\left({\bar A_0\over A_0}\right).
\eeq
Since, by definition, only one strong channel contributes to $\lambda_0$,
there is indeed no CP violation in decay in (\ref{defepsth}).
It is simple to show that $\Re\ \epsK\neq0$ is a manifestation
of CP violation in mixing while $\Im\ \epsK\neq0$ is a manifestation
of CP violation in the interference between decays with and without 
mixing. Since experimentally $\arg\epsK\approx\pi/4$, the two
contributions are comparable. It is also clear that $\epsK\neq0$
is a manifestation of indirect CP violation: it could be described
entirely in terms of a CP violating phase in the $M_{12}$ amplitude.

The experimental definition of the $\epspK$ parameter is 
\beq\label{defepspex}
\epspK\equiv{1\over3}(\eta_{+-}-\eta_{00}).
\eeq
The theoretical expression is
\beq\label{defepspth}
\epspK\approx{1\over6}(\lambda_{00}-\lambda_{+-}).
\eeq
Obviously, any type of CP violation which is independent of the
final state does not contribute to $\epspK$. Consequently,
there is no contribution from CP violation in mixing to (\ref{defepspth}). 
It is simple to show that $\Re\ \epspK\neq0$ 
is a manifestation of CP violation in decay while $\Im\ \epspK\neq0$ 
is a manifestation of CP violation in the interference between decays 
with and without mixing. Following our explanations in the previous
section, we learn that $\epspK\neq0$ is a manifestation of direct 
CP violation: it requires $\phi_2-\phi_0\neq0$ (where $\phi_I$ is
the CP violating phase in the $A_I$ amplitude defined in (\ref{defAI})).

\section{CP VIOLATION IN THE STANDARD  MODEL}
\subsection{Introduction}
Within the Standard Model, CP violation can only arise from the
Yukawa interactions: 
\beq\label{Hint}
-{\cal L}_Y=Y^d_{ij}{\overline {Q^I_{Li}}}\phi d^I_{Rj}
+Y^u_{ij}{\overline {Q^I_{Li}}}\tilde\phi u^I_{Rj}
+Y^\ell_{ij}{\overline {L^I_{Li}}}\phi\ell^I_{Rj}.
\eeq
The various fermion representations of $SU(3)_{\rm C}\times SU(2)_{\rm L}
\times U(1)_{\rm Y}$ are denoted here by
\beq\label{MSSMrep}
Q_i^I(3,2)_{+1/6},\ \ \bar u_i^I(\bar3,1)_{-2/3},\ \ 
\bar d_i^I(\bar3,1)_{+1/3},\ \ L_i^I(1,2)_{-1/2},\ \ 
\bar\ell_i^I(1,1)_{+1},
\eeq
and the Higgs representation is $\phi(1,2)_{+1/2}$ 
($\tilde\phi=i\sigma_2\phi^*$). There are 27 complex parameters in
the three Yukawa matrices, but not all of them are physical. If the
Yukawa couplings are switched off, the Standard Model has 
(in addition to a discrete CP symmetry) a global $U(3)^5$ symmetry. 
But the subgroup of $U(1)_B\times U(1)_e\times
U(1)_\mu\times U(1)_\tau$ remains a symmetry of the Standard Model
even in the presence of non-zero Yukawa couplings. Since a $3\times3$
unitary matrix has three real and six imaginary parameters, we conclude
that $15$ real and $26$ imaginary parameters in the Yukawa matrices are 
not physical, leaving twelve real and one imaginary physical parameters.

It is easy to identify the physical parameters in the mass basis.
Nine of the real parameters are the charged fermion masses.
All other parameters are related to the CKM matrix $V_{\rm CKM}$, 
that is the quark mixing matrix that parametrizes the charged gauge
boson interactions \cite{Cabi,KoMa}:
\beq\label{Wmas}
-{\cal L}_{W^\pm}={g\over\sqrt2}{\overline {u_{Li}}}\gamma^\mu
(V_{\rm CKM})_{ij}d_{Lj} W_\mu^++{\rm h.c.}.
\eeq
The unitary $V_{\rm CKM}$ can be parametrized with three real 
mixing angles and a single phase. The single irremovable 
phase in the CKM matrix is the only source CP violation within the
Standard Model.

In the Wolfenstein parametrization of $V_{\rm CKM}$, the four mixing
parameters are $(\lambda,A,\rho,\eta)$ with
$\lambda=|V_{us}|=0.22$ playing the role of an expansion parameter
and $\eta$ representing the CP violating phase \cite{WOLpar}:
\beq\label{WCKM}
V=\pmatrix{1-{\lambda^2\over2}&\lambda&A\lambda^3(\rho-i\eta)\cr
-\lambda&1-{\lambda^2\over2}&A\lambda^2\cr
A\lambda^3(1-\rho-i\eta)&-A\lambda^2&1\cr}+\O(\lambda^4).
\eeq

The fact that there is a single CP violating parameter in the SM
can be seen also in another useful way. The unitarity of the CKM
matrix leads to various relations among the matrix elements, {\it e.g.}
\beq\label{Unitds}
V_{ud}V_{us}^*+V_{cd}V_{cs}^*+V_{td}V_{ts}^*=0,
\eeq
\beq\label{Unitsb}
V_{us}V_{ub}^*+V_{cs}V_{cb}^*+V_{ts}V_{tb}^*=0,
\eeq
\beq\label{Unitdb}
V_{ud}V_{ub}^*+V_{cd}V_{cb}^*+V_{td}V_{tb}^*=0.
\eeq
Each of the three relations (\ref{Unitds})$-$(\ref{Unitdb}) requires 
the sum of three complex quantities to vanish and so can be geometrically
represented in the complex plane as a triangle. These are
``the unitarity triangles", though the term ``unitarity triangle"
is usually reserved for the relation (\ref{Unitdb}) only. It is a surprising
feature of the CKM matrix that all unitarity triangles are equal in area.
For any choice of $i,j,k,l=1,2,3$, one can define a quantity $J$
according to \cite{Jarl}
\beq\label{defJ}
\Im[V_{ij}V_{kl}V_{il}^*V_{kj}^*]=J\sum_{m,n=1}^3\epsilon_{ikm}\epsilon_{jln}.
\eeq
Then, the area of each unitarity triangle equals $|J|/2$ while
the sign of $J$ gives the direction of the complex vectors
around the triangles. CP is violated in the Standard Model only if 
$J\neq0$. The quantity $J$ can then be taken as the CP violating
parameter of the SM. The area of the triangles
is then related to the size of the Standard Model CP violation.
The relation between Jarlskog's measure of CP violation $J$
and the Wolfenstein parameters $(\lambda,A,\eta)$ is given by
\beq\label{JAre}
J\simeq \lambda^6 A^2\eta.
\eeq
 
The rescaled unitarity triangle  is derived from (\ref{Unitdb})
by (a) choosing a phase convention such that $(V_{cd}V_{cb}^*)$
is real, and (b) dividing the lengths of all sides by $|V_{cd}V_{cb}^*|$.
Step (a) aligns one side of the triangle with the real axis, and
step (b) makes the length of this side 1. The form of the triangle
is unchanged. Two vertices of the rescaled unitarity triangle are
thus fixed at (0,0) and (1,0). The coordinates of the remaining
vertex correspond to the Wolfenstein parameters $(\rho,\eta)$.

Depicting the rescaled unitarity triangle in the
$(\rho,\eta)$ plane, the lengths of the two complex sides are
\beq\label{RbRt}
R_u\equiv\sqrt{\rho^2+\eta^2}={1\over\lambda}
\left|{V_{ub}\over V_{cb}}\right|,\ \ \
R_t\equiv\sqrt{(1-\rho)^2+\eta^2}={1\over\lambda}
\left|{V_{td}\over V_{cb}}\right|.
\eeq
The three angles of the  unitarity triangle are denoted
by $\alpha,\beta$ and $\gamma$ \cite{DDGN}:
\beq\label{abcangles}
\alpha\equiv\arg\left[-{V_{td}V_{tb}^*\over V_{ud}V_{ub}^*}\right],\ \ \
\beta\equiv\arg\left[-{V_{cd}V_{cb}^*\over V_{td}V_{tb}^*}\right],\ \ \
\gamma\equiv\arg\left[-{V_{ud}V_{ub}^*\over V_{cd}V_{cb}^*}\right].
\eeq
They are physical quantities and, we will soon see, can be
independently measured by CP asymmetries in $B$ decays.

To make predictions for future measurements of CP violating observables,
we need to find the allowed ranges for the CKM phases. There are three
ways to determine the CKM parameters (see {\it e.g.} \cite{HarNir}):
\begin{itemize}
\item[(i)] {\bf Direct measurements} are related to SM tree level 
processes. At present,
we have direct measurements of $|V_{ud}|$, $|V_{us}|$, $|V_{ub}|$,
$|V_{cd}|$, $|V_{cs}|$, $|V_{cb}|$ and $|V_{tb}|$. 
\item[(ii)] {\bf CKM Unitarity}  ($V_{\rm CKM}^\dagger
V_{\rm CKM}={\bf 1}$) relates the various matrix elements. At present, 
these relations are useful to constrain
$|V_{td}|$, $|V_{ts}|$, $|V_{tb}|$ and $|V_{cs}|$. 
\item[(iii)] {\bf Indirect measurements} are related to SM loop processes. 
At present, we constrain in this way $|V_{tb}V_{td}|$ (from $\Delta m_B$ 
and  $\Delta m_{B_s}$) and $\delta_{\rm KM}$ (from $\epsK$).
\end{itemize}

When all available data is taken into account, we find 
\cite{NirCern,PlSc,Plas,GNPS,BaBar}:
\beq\label{recons}
-0.15\leq\rho\leq+0.35,\ \ \ +0.20\leq\eta\leq+0.45,
\eeq
\beq\label{abcons}
0.4\leq\sin2\beta\leq0.8,\ \ \ -0.9\leq\sin2\alpha\leq1.0,\ \ \ 
0.23\leq\sin^2\gamma\leq1.0.
\eeq
Of course, there are correlations between the various parameters.
The full information in the $(\rho,\eta)$ and in the $(\sin2\alpha,
\sin2\beta)$ plane is given in fig. 1.

\begin{figure}
\centerline{$(a)$}
\centerline{
\psfig{file=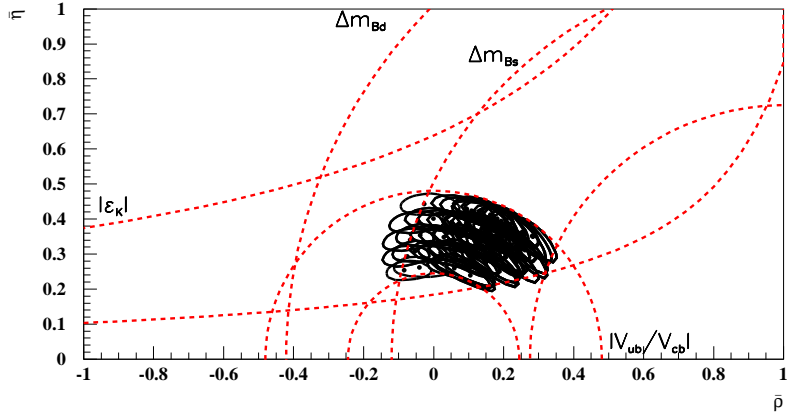,width=370pt,bbllx=0pt,bblly=410pt,bburx=612pt,bbury=740pt
}}
\centerline{$(b)$}
\centerline{
\psfig{file=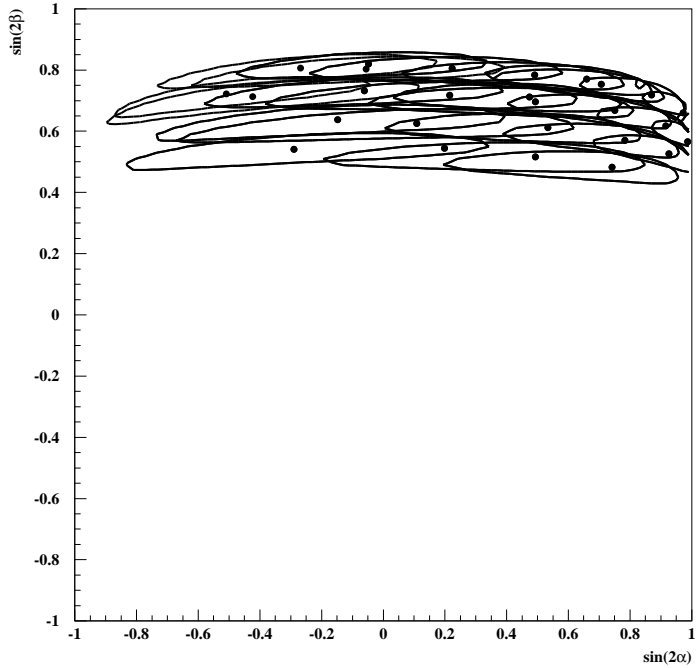,width=320pt,bbllx=0pt,bblly=160pt,bburx=612pt,bbury=653pt
}}
\caption{The present allowed range $(a)$ in the $\rho-\eta$ plane
and $(b)$ in the $\sin2\alpha-\sin2\beta$ plane
using constraints from $|V_{cb}|$,
$|V_{ub}/V_{cb}|$, $\Delta m_{B_d}$, $\epsK$ and $\Delta m_{B_s}$.
For the methods and the data used in this analysis, see ref. [45].
\label{presentRE}}
\end{figure}

Since the Standard Model contains only a single independent
CP-violating phase, all possible CP-violating effects in this
theory are very closely related. Consequently, the pattern of
CP-violations in $B$ decays is strongly constrained. The goal
of $B$ factories is to test whether this pattern occurs in Nature.
If no new physics is discovered, we will have a much improved
determionation of the CKM parameters. The impact of measurements
of CP asymmetries in $B$ decays and of the rate of $K_L\ra\pi\nu\nu$
is demonstrated in fig. 2.

\begin{figure}
\centerline{$(a)$}
\centerline{
\psfig{file=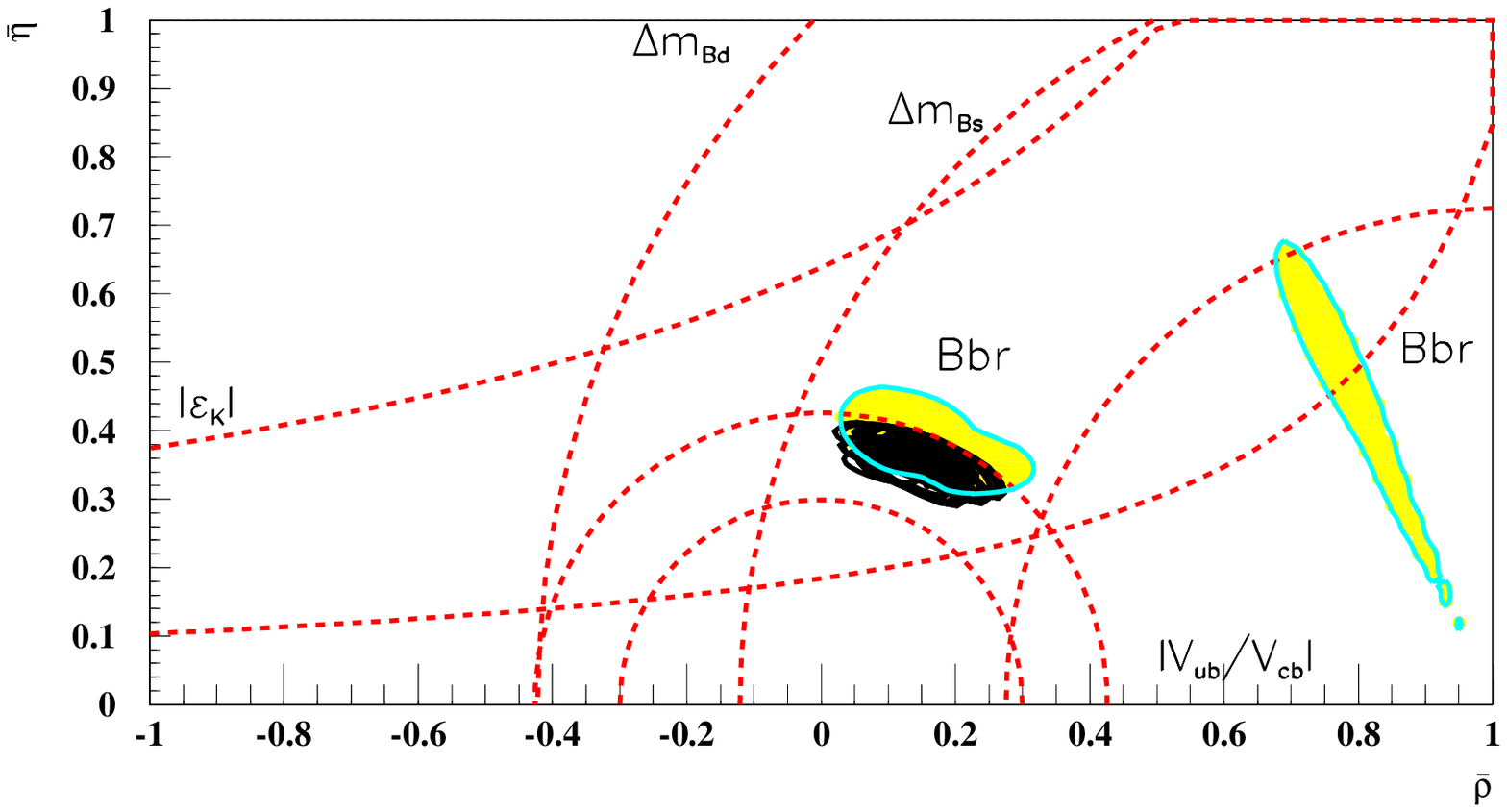,width=370pt,bbllx=0pt,bblly=410pt,bburx=612pt,bbury=700pt
}}
\centerline{$(b)$}
\centerline{
\psfig{file=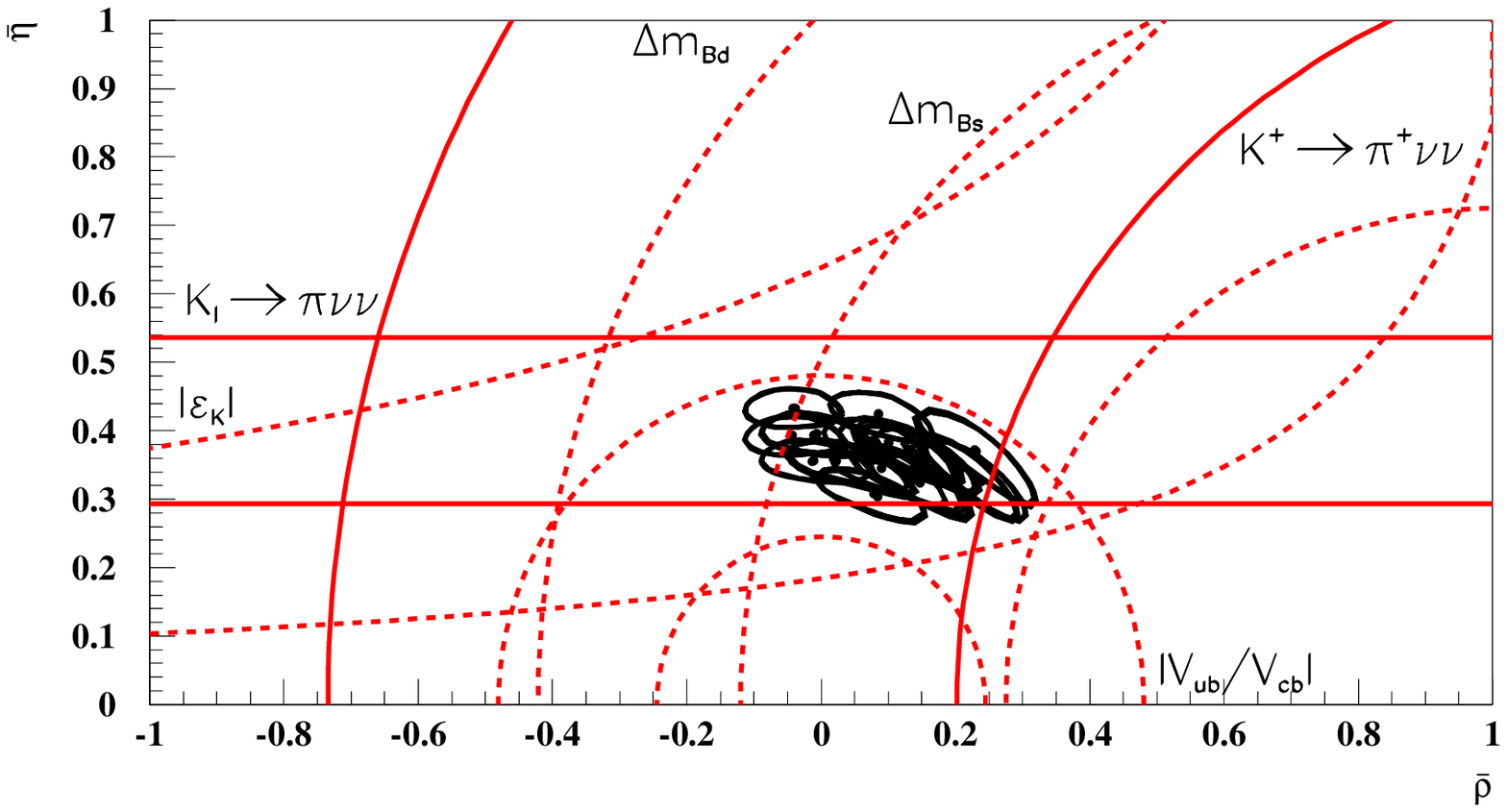,width=370pt,bbllx=0pt,bblly=260pt,bburx=612pt,bbury=700pt
}}
\caption{(a) The effect of the CP asymmetries in $B$ decays
on the constraints in the $\rho-\eta$ plane. We use hypothetical 
ranges appropriate for 180 fb$^{-1}$ integrated luminosity
in the $B$-factories. 
(b) The effect of the $K\ra\pi\nu\bar\nu$ measurements
on the constraints in the $\rho-\eta$ plane.
We use the hypothetical ranges 
BR($K^+\ra\pi^+\nu\bar\nu)=(1.0\pm0.1)\times10^{-10}$,
BR($K_L\ra\pi^0\nu\bar\nu)=(3.0\pm0.3)\times10^{-11}$.  
For the methods and the data used in this analysis, see ref. [45].
\label{kpnnRE}}
\end{figure}

\subsection{CP Violation in Semileptonic Decays of Neutral $B$ Mesons}
In the $B_d$ system we expect model independently that $\Gamma_{12}\ll 
M_{12}$. Moreover, within the SM and assuming that the box diagram 
(with a cut) is appropriate to estimate $\Gamma_{12}$, we can actually 
calculate the two quantities from quark diagrams \cite{BKUS}.
The calculation gives 
\beq\label{SMGtoM}
{\Gamma_{12}\over M_{12}}=-{3\pi\over2}{1\over f_2(m_t^2/m_W^2)}
{m_b^2\over m_t^2}\left(1+{8\over3}{m_c^2\over m_b^2}
{V_{cb}V_{cd}^*\over V_{tb}V_{td}^*}\right).
\eeq
This confirms our order of magnitude estimate, $|\Gamma_{12}/M_{12}|
\lsim10^{-2}$. The deviation of $|q/p|$ from unity is proportional
to $\Im(\Gamma_{12}/M_{12})$ which is even further suppressed by
another order of magnitude:
\beq\label{AbsqpSM}
1-\left|{q\over p}\right|={1\over2}\Im{\Gamma_{12}\over M_{12}}=
{4\pi\over f_2(m_t^2/m_W^2)}{m_c^2\over m_t^2}{J\over|V_{tb}V_{td}^*|^2}
\sim10^{-3}.
\eeq
Note that the suppression comes from the $(m_c^2/m_t^2)$ factor.
The last term is the ratio of the area of the unitarity triangle to
the length of one of its sides squared, so it is $\O(1)$. In contrast,
for the $B_s$ system, where (\ref{AbsqpSM}) holds except that $V_{td}$ is
replaced by $V_{ts}$, there is an additional suppression from
$J/|V_{tb}V_{ts}^*|^2\sim10^{-2}$ (see the corresponding unitarity triangle).

The above estimate of CP violation in mixing suffers from large 
uncertainties (of order 30\% \cite{Alek}\ or even higher \cite{WolGot})
related to the use of a quark diagram to describe $\Gamma_{12}$ .

We remind the reader that (\ref{AbsqpSM}) gives the leading contribution
to the CP asymmetry in semileptonic decays:
\beq\label{SMSLB}
a_{\rm SL}\approx2(1-|q/p|).
\eeq
The smallness of the predicted asymmetry will make its measurement
a rather challenging task.

\subsection{CP Violation in Hadronic Decays of Neutral $B$ Mesons}
In the previous subsection we estimated the effect of CP violation
in mixing to be of $\O(10^{-3})$ within the Standard Model, and
$\leq\O(|\Gamma_{12}/M_{12}|)\sim10^{-2}$ model independently 
(for a recent discussion, see \cite{RaSu}).
In semileptonic decays, CP violation in mixing is the leading effect
and therefore it can be measured through $a_{\rm SL}$. In purely hadronic
$B$ decays, however, CP violation in decay and in the interference of
decays with and without mixing is $\geq \O(10^{-2})$. We can therefore
safely neglect CP violation in mixing in the following discussion and use
\beq\label{qpSM}
{q\over p}=-{M_{12}^*\over|M_{12}|}=
{V_{tb}^*V_{td}\over V_{tb}V_{td}^*}\omega_B.
\eeq
(From here on we omit the convention-dependent quark phases $\omega_q$
defined in eq. (\ref{CPofq}). Our final expressions for physical
quantities are of course unaffected by such omission.)

A crucial question is then whether CP violation in decay is comparable to 
the CP violation in the interference of decays with and without mixing or 
negligible.  In the first case, we can use the corresponding charged $B$ 
decays to observe effects of CP violation in decay. In the latter case, 
CP asymmetries in neutral $B$ decays are subject to clean theoretical 
interpretation: we will either have precise measurements of CKM parameters 
or be provided with unambiguous evidence for new physics. The question of 
the relative size of CP violation in decay can only be answered
on a channel by channel basis, which is what we do in this section.

Most channels have contributions from both tree- and three types of
penguin-diagrams, the latter classified according to the identity of the 
quark in the loop, as diagrams with different intermediate quarks may have 
both different strong phases and different weak phases \cite{BSS}.
On the other hand, the subdivision of tree processes into
spectator, exchange and annihilation diagrams is unimportant in this
respect since they all carry the same weak phase.

While quark diagrams can be easily classified in this way, the
description of $B$ decays is not so neatly divided into tree and penguin
contributions once long distance physics effects are taken into
account. Rescattering processes can change the quark content of the final
state and confuse the identification of a contribution. There is no
physical distinction between rescattered tree diagrams and long-distance
contributions to the cuts of a penguin diagram.  While these issues
complicate estimates of various rates, they can always be avoided in
describing the weak phase structure of $B$-decay amplitudes. The decay
amplitudes for $b\rightarrow q \bar q q^\prime$ can always be
written as a sum of three terms with definite CKM coefficients:
\beq\label{defCKM}
A(q \bar q q^\prime)=  V_{tb}V^*_{tq^\prime}P^t_{q^\prime} +
V_{cb}V_{cq^\prime}^*(T_{c\bar c q^\prime}\delta_{qc} + P^c_{q^\prime})+
V_{ub}V_{uq^\prime}^*(T_{u \bar u q^\prime}\delta_{qu} + P^u_{q^\prime}).
\eeq
Here $P $ and $T$ denote contributions from tree and penguin diagrams,
excluding the CKM factors. As they stand, the $P$ terms are not
well defined because of the divergences of the penguin diagrams.  Only
differences of penguin diagrams are finite and well defined. However
already we see that diagrams that can be mixed by rescattering effects
always appear with the same CKM coefficients and hence that a separation
of these terms is not needed when discussing weak phase structure. Now it is 
useful to use eqs. (\ref{Unitsb}) and (\ref{Unitdb}) to eliminate one of the 
three terms, by writing its CKM coefficient as minus the sum of the other two.

In the case of $q \bar q s$ decays it is convenient to remove the
$V_{tb} V^*_{ts}$ term. Then
\beq\label{ccstype}
A(c\bar c s)= V_{cb}V_{cs}^*(T_{c\bar cs}+P^c_s-P^t_s)
+V_{ub}V^*_{us}(P^u_s-P^t_s), 
\eeq
\beq\label{uustype}
A(u\bar u s)= V_{cb}V_{cs}^*(P^c_s-P^t_s)
+V_{ub}V^*_{us}(T_{u\bar us}+P^u_s-P^t_s), 
\eeq
\beq\label{ssstype}
A(s\bar s s)= V_{cb}V_{cs}^*(P^c_s-P^t_s)+V_{ub}V^*_{us}(P^u_s- P^t_s).
\eeq
In these expressions only differences of penguin contributions occur,
which makes the cancellation of the ultraviolet divergences of these
diagrams explicit. Furthermore, the second term has a CKM coefficient that
is much smaller, by $\O(\lambda^2)$, than the first. Hence this grouping
is useful in classifying the expected CP violation in decay.  

In the case of $q \bar q d $ decays the three CKM coefficients are
of similar magnitude. The convention is then to retain the
$V_{tb}V^*_{td}$ term because, in the Standard Model, the phase
difference between this weak phase and half the mixing weak phase is
zero. Thus only one unknown weak phase enters the calculation of the
interference between decays with and without mixing.
We can choose to eliminate which of the other terms does not
have a tree contribution. In the cases $q=s$ or $d$, since neither has a
tree contribution either term can be removed. Thus we write
\beq\label{ccdtype}
A(c\bar c d)=V_{tb}V_{td}^*(P^t_d-P^u_d)
+ V_{cb}V_{cd}^*(T_{c\bar c d}+P^c_d-P^u_d), 
\eeq
\beq\label{uudtype}
A(u\bar u d)=V_{tb}V_{td}^*(P^t_d-P^c_d)
 +V_{ub}V_{ud}^*(T_{u\bar u d}+P^u_d - P^c_d), 
\eeq
\beq\label{ssdtype}
A(s\bar s d)=V_{tb}V_{td}^*(P^t_d-P^u_d)+V_{cb}V_{cd}^*(P^c_d-P^u_d).
\eeq
Again only differences of penguin amplitudes occur. Furthermore the
difference of penguin terms that occurs in the second term would vanish
if the charm and up quark masses were equal, and thus is GIM suppressed
\cite{GIM}. However, even in modes with no tree contribution, $(s\bar s d)$,
the interference of the two terms can still give significant CP violation.

The penguin processes all involve the emission of a neutral boson,
either a gluon (strong penguins) or a photon or $Z$ boson (electroweak
penguins). Excluding the CKM coefficients, the ratio of the contribution
from the difference between a top and light quark strong penguin diagram
to the contribution from a tree diagram is of order
\beq\label{pengtree}
r_{PT} = { P^t-P^{light}\over T_{q \bar q q^\prime
 }}\approx{\alpha_s\over12\pi}\ln{m_t^2\over m_b^2}.
\eeq
This is a factor of $\O(0.03)$. However this estimate does not include
the effect of hadronic matrix elements, which are the probability
factor to produce a particular final state particle content from a
particular quark content. Since this probability differs for different
kinematics, color flow and spin structures, it can be different for tree
and penguin contributions and may partially compensate the coupling
constant suppression of the penguin term. Recent CLEO results on
$BR(B\ra K\pi)$ and $BR(B\ra\pi\pi)$ \cite{CLEOpen}\
suggest that the matrix element of penguin operators is indeed enhanced
compared to that of tree operators. The enhancement could be by a factor
of a few, leading to
\beq\label{pentre}
r_{PT}\sim\lambda^2-\lambda.
\eeq
(Note that $r_{PT}$ does not depend on the CKM parameters.
We use powers of the Wolfenstein parameter $\lambda$ to
quantify our estimate for $r_{PT}$ is order to simplify
the comparison between the size of CP violation in decay
and CP violation in the interference between decays with
and without mixing.) Electroweak penguin difference
terms are even more suppressed since they have an $\alpha_{\rm EM}$ or
$\alpha_W$ instead of the $\alpha_s$ factor in (\ref{pengtree}), but certain
$Z$-contributions are enhanced by the large top
quark mass and so can be non-negligible.

We thus classify $B$ decays into four classes. Classes (i) and (ii) are
expected to have relatively small CP violation in decay and hence are
particularly interesting for extracting CKM parameters from
interference of decays with and without mixing. In the remaining
two classes, CP violation in decay could be significant and the neutral
decay asymmetries cannot be cleanly interpreted in terms of CKM phases.

\begin{itemize}
\item[(i)] Decays dominated by a single term: $b\ra c\bar cs$ and $b\ra
s\bar ss$. The Standard Model cleanly predicts very small
CP violation in decay: $\O(\lambda^4-\lambda^3)$ for $b\ra c\bar cs$
and $\O(\lambda^2)$ for $b\ra s\bar ss$.
Any observation of large CP asymmetries in charged $B$ decays
for these channels would be a clue to physics beyond the Standard Model.
The corresponding neutral modes have cleanly predicted relationships
between CKM parameters and the measured asymmetry from interference
between decays with and without mixing. The modes
$B\ra\psi K$ and $B\ra\phi K$ are examples of this class.
\item[(ii)] Decays with a small second term: $b\ra c\bar cd$ and $b\ra
u \bar ud$. The expectation that penguin-only contributions are supressed
compared to tree contributions suggests that these modes will have small
effects of CP violation in decay, of $\O(\lambda^2-\lambda)$, and an
approximate prediction for the relationship between measured asymmetries
in neutral decays and CKM
phases can be made. Examples here are $B\ra\ DD$ and $B\ra\pi\pi$.
\item[(iii)] Decays with a suppressed tree contribution: $b\ra u \bar us$.
The tree amplitude is suppressed by small mixing angles, $V_{ub}V_{us}$.
The no-tree term may be comparable or even dominate and give large
interference effects. An example is $B\ra\rho K$.
\item[(iv)] Decays with no tree contribution: $b\ra s\bar s d$. Here the
interference comes from penguin contributions with different charge 2/3
quarks in the loop and gives CP violation in decay that could be as large
as 10\% \cite{Flei,BuFl}. An example is $B\ra KK$.
\end{itemize}

Note that if the penguin enhancement is significant,
then some of the decay modes listed in class (ii) might actually fit
better in class (iii). For example, it is possible that $b\ra u\bar ud$
decays have comparable contributions from tree and penguin amplitudes.
On the other hand, this would also mean that some modes listed in class
(iii) could be dominated by a single penguin term. For such cases
an approximate relationship between measured
asymmetries in neutral decays and CKM phases can be made.

\subsection{CP Violation in the Interference Between $B$ Decays With and
Without Mixing}
Let us first discuss an example of class (i), $B\ra\psi K_S$. A new
ingredient in the analysis is the effect of $K-\bar K$ mixing. For decays
with a single $K_S$ in the final state, $K-\bar K$ mixing is essential
because $B^0\ra K^0$ and $\bar B^0\ra\bar K^0$, and interference is
possible only due to $K-\bar K$ mixing. This adds a factor of
\beq\label{qpK}
\left({p\over q}\right)_K={V_{cs}V_{cd}^*\over V_{cs}^*V_{cd}}\omega_K^*
\eeq
into $(\bar A/A)$. The quark subprocess in $\bar B^0\ra\psi\bar K^0$ is
$b\ra c\bar cs$ which is dominated by the $W$-mediated tree diagram:
\beq\label{ApsiK}
{\bar A_{\psi K_S}\over A_{\psi K_S}}= \eta_{\psi K_S}
\left({V_{cb}V_{cs}^*\over V_{cb}^*V_{cs}}\right)
\left({V_{cs}V_{cd}^*\over V_{cs}^*V_{cd}}\right)\omega_B^*.
\eeq
The CP-eigenvalue of the state is $\eta_{\psi K_S} = -1$.
Combining (\ref{qpSM}) and (\ref{ApsiK}), we find
\beq\label{lampsiK}
\lambda(B\ra\psi K_S)=-\left({V_{tb}^*V_{td}\over
V_{tb}V_{td}^*}\right)\left({V_{cb}V_{cs}^*\over V_{cb}^*V_{cs}}\right)
\left({V_{cd}^*V_{cs}\over V_{cd}V_{cs}^*}\right)
\ \Longrightarrow\ \Im\lambda_{\psi K_S}=\sin(2\beta).
\eeq

The second term in (\ref{ccstype}) is of order $\lambda^2r_{PT}$
for this decay and thus eq. (\ref{lampsiK}) is clean of hadronic
uncertainties to $\O(10^{-3})$. Consequently, this measurement can give
the theoretically cleanest determination of a CKM parameter, even
cleaner than the determination of $|V_{us}|$ from $K\ra\pi\ell\nu$. (If
BR($K_L\ra\pi\nu\bar\nu$) is measured, it will give a comparably clean
determination of $\eta$.)

A second example of a theoretically clean mode in class (i) is
$B\ra\phi K_S$. The quark subprocess involves FCNC
and cannot proceed via a tree level SM diagram. The leading
contribution comes from penguin diagrams. The two terms in eq.
(\ref{ssstype}) are now both differences of penguins but the second term
is CKM suppressed and thus of $\O(\lambda^2)$ compared to the first. Thus
CP violation in the decay is at most a few percent, and can be neglected 
in the analysis of asymmetries in this channel.  The analysis is similar to 
the $\psi K_S$ case, and the asymmetry is proportional to $\sin(2\beta)$.

The same quark subprocesses give theoretically clean CP
asymmetries also in $B_s$ decays. These asymmetries are, however,
very small since the relative phase between the mixing amplitude
and the decay amplitudes ($\beta_s$ defined below) is very small.

The best known example of class (ii) is $B\ra\pi\pi$. The quark subprocess
is $b\ra u\bar ud$ which is dominated by the $W$-mediated tree diagram.
Neglecting for the moment the second, pure penguin, term in eq.
(\ref{uudtype}) we find
\beq\label{Apipi}
{\bar A_{\pi\pi}\over A_{\pi\pi}}=\eta_{\pi\pi}{V_{ub}V_{ud}^*\over
 V_{ub}^*V_{ud}}\omega_B^*.
\eeq
The CP eigenvalue for two pions is $+1$.
Combining (\ref{qpSM}) and (\ref{Apipi}), we get
\beq\label{lampipi}
\lambda(B\ra\pi^+\pi^-)=\left({V_{tb}^*V_{td}\over V_{tb}V_{td}^*}\right)
\left({V_{ud}^*V_{ub}\over V_{ud}V_{ub}^*}\right)
\ \Longrightarrow\ \Im\lambda_{\pi\pi}=\sin(2\alpha).
\eeq
The pure penguin term in eq. (\ref{uudtype}) has a weak phase,
$\arg(V_{td}^*V_{tb})$, different from the term with the tree
contribution, so it modifies both $\Im\lambda_{\pi\pi}$ and (if there are
non-trivial strong phases) $|\lambda_{\pi\pi}|$. The recent CLEO results
mentioned above suggest that the penguin contribution
to $B\ra\pi\pi$ channel is significant, probably  $10\%$ or more. This
then introduces CP violation in decay, unless the strong phases
cancel (or are zero, as suggested by factorization arguments).
The resulting hadronic uncertainty can be eliminated using isospin analysis
\cite{GrLo}. This requires a measurement of the rates for the
isospin-related channels $B^+ \ra \pi^+ \pi^0$ and $B^0 \ra \pi^0\pi^0$
as well as the corresponding CP-conjugate processes. The rate for
$\pi^0\pi^0$ is expected to be small and the measurement is difficult,
but even an upper bound on this rate can be used to limit the magnitude
of hadronic uncertainties \cite{GrQu}.

Related but slightly more complicated channels with the same underlying
quark structure are $B\ra\rho^0\pi^0$ and $B\ra a_1^0\pi^0$. Again an
analysis involving the isospin-related channels can be used to help
eliminate hadronic uncertainties from CP violations in the decays
\cite{LNQS,SnQu}.

Channels such as $\rho\rho$ and $a_1\rho$ could in principle also be
studied, using angular analysis to determine the mixture of
CP-even and CP-odd contributions.

The analysis of $B\ra D^+D^-$ proceeds along very similar lines. The quark 
subprocess here is $b\ra c\bar cd$, and so the tree contribution gives
\beq\label{lamDD}
\lambda(B\ra D^+D^-)=
\eta_{D^+D^-}\left({V_{tb}^*V_{td}\over V_{tb}V_{td}^*}\right)
\left({V_{cd}^*V_{cb}\over V_{cd}V_{cb}^*}\right)
\ \Longrightarrow\ \Im\lambda_{DD}=-\sin(2\beta),
\eeq
where we used $\eta_{D^+D^-}=+1$. Again, there are hadronic uncertainties 
due to the pure penguin term in (\ref{ccdtype}), but they are estimated to 
be small.

In all cases the above discussions have neglected the distinction
between strong penguins and electroweak penguins. The CKM phase
structure of both types of penguins is the same. The only place where
this distinction becomes important is when an isospin argument is used
to remove hadronic uncertainties due to penguin contributions. These
arguments are based on the fact that gluons have isospin zero, and
hence strong penguin processes have definite $\Delta I$. Photons and
$Z$-bosons on the other hand contribute to more than one $\Delta I$
transition and hence cannot be separated from tree terms by isospin
analysis. In most cases electroweak penguins are small, typically no more
than ten percent of the corresponding strong penguins and so their
effects can safely be neglected. However in cases (iii) and (iv), where tree
contributions are small or absent, their effects may need to be considered.
(A full review of the role of electroweak penguins in $B$
decays has been given in ref. \cite{ewpenreview}.)

\subsection{Unitarity Triangles}
One can obtain an intuitive understanding of the Standard Model
CP violation in the interference between decays with and without mixing
by examining the unitarity triangles. It is instructive to draw the
three triangles, (\ref{Unitds}), (\ref{Unitsb}) and (\ref{Unitdb}), knowing 
the experimental values (within errors) for the various $|V_{ij}|$. In the 
first triangle (\ref{Unitds}), one side is of $\O(\lambda^5)$ and therefore
much shorter than the other, $\O(\lambda)$, sides. In the second
triangle (\ref{Unitsb}), one side is of $\O(\lambda^4)$ and therefore shorter
than the other, $\O(\lambda^2)$, sides. In the third triangle (\ref{Unitdb}),
all sides have lengths of $\O(\lambda^3)$. The first two triangles then
almost collapse to a line while the third one is open.

Let us examine the CP asymmetries in the leading decays into
final CP eigenstates. For the $B$ mesons, the size of these
asymmetries ({\it e.g.} $\Im\lambda_{\psi K_S}$) depends on
$\beta$ because it gives the difference between half the phase of
the $B-\bar B$ mixing amplitude and the phase of the decay
amplitudes. The form of the third unitarity triangle, (\ref{Unitdb}),
implies that $\beta=\O(1)$, which explains
why these asymmetries are expected to be large.

It is useful to define the analog phases for the
$B_s$ meson, $\beta_s$, and the $K$ meson, $\beta_K$:
\beq\label{bbangles}
\beta_s\equiv\arg\left[-{V_{ts}V_{tb}^*\over V_{cs}V_{cb}^*}\right],\ \ \
\beta_K\equiv\arg\left[-{V_{cs}V_{cd}^*\over V_{us}V_{ud}^*}\right].
\eeq
The angles $\beta_s$ and $\beta_K$ can be seen to be the small angles of
the second and first unitarity triangles, (\ref{Unitsb}) and 
(\ref{Unitds}), respectively.
This gives an intuitive understanding of why CP violation is small in the
leading $K$ decays (that is $\epsK$ measured in $K\rightarrow\pi\pi$
decays) and is expected to be small in the leading $B_s$ decays ({\it e.g.}
$B_s\rightarrow\psi\phi$). Decays related to the short sides of these
triangles are rare but could exhibit
significant CP violation. Actually, the large angles in the (\ref{Unitds})
triangle are approximately $\beta$ and $\pi-\beta$, which explains why
CP violation in $K\ra\pi\nu\bar\nu$ is related to $\beta$ and expected
to be large. The large angles in the (\ref{Unitsb})
triangle are approximately $\gamma$ and $\pi-\gamma$. This explains
why the CP asymmetry in $B_s\ra\rho K_S$ is related to $\gamma$
and expected to be large. (Note, however, that this mode gets
comparable contributions from penguin and tree diagrams and does
not give a clean CKM measurement \cite{BuFl}.)

\section{CP VIOLATION BEYOND THE STANDARD MODEL}

The Standard Model picture of CP violation is rather unique and highly
predictive. In particular, we would like to point out the following features:
\begin{itemize}
\item[(i)] CP is broken explicitly.
\item[(ii)] All CP violation arises from a single phase, that is $\dKM$.
\item[(iii)] The measured value of $\epsK$ requires that $\dKM$
is of order one. (In other words, CP is not an approximate symmetry of
the Standard Model.)
\item[(iv)] The values of all other CP violating observables
can be predicted. In particular, CP violation in $B\ra\psi K_S$
(and similarly various other CP asymmetries in $B$ decays),
and in $K\ra\pi\nu\bar\nu$ are expected to be of order one.
\end{itemize}

The commonly repeated statement that CP violation is one of the least
tested aspects of the Standard Model is well demonstrated by the fact
that none of the above features necessarily holds in the presence
of New Physics. In particular, there are viable models of new physics
({\it e.g.} certain supersymmetric models) with the following features:
\begin{itemize}
\item[(i)] CP is broken spontaneously.
\item[(ii)] There are many CP violating phases (even in the low
energy effective theory).
\item[(iii)]  CP is an approximate symmetry, with all CP violating
phases small (usually $10^{-3}\lsim\phi_{\rm CP}\lsim 10^{-2}$).
\item[(iv)] Values of CP violating observables
can be predicted and could be very different from the Standard
Model predictions (except, of course, $\epsK$). In particular,
$\Im\lambda_{\psi K_S}$ and $a_{\pi\nu\bar\nu}$ could both be $\ll1$.
\end{itemize}

We emphasize that various extensions of the SM modify its predictions
of CP violation in various ways. The models of approximate CP described
above are only one example of how the predictions can be dramatically
violated. 

To understand how the Standard Model predictions could be modified
by New Physics, we will focus on CP violation in the interference
between decays with and without mixing. As explained above, it is
this type of CP violation which, due to its theoretical cleanliness,
may give unambiguous evidence for New Physics most easily.

\subsection{CP Violation as a Probe of Flavor Beyond the Standard Model}
Let us consider five specific CP violating observables.
\begin{itemize}
\item[(i)] $\Im\lambda_{\psi K_S}$, the CP asymmetry in $B\rightarrow\psi K_S$.
This measurement will cleanly determine the relative phase between
the $B-\bar B$ mixing amplitude and the $b\ra c\bar cs$ decay amplitude
($\sin2\beta$ in the Standard Model).
The $b\ra c\bar cs$ decay has Standard Model tree contributions and
therefore is very unlikely to be significantly affected by new physics.
On the other hand, the mixing amplitude can be easily modified
by new physics. We parametrize such a modification by a phase $\theta_d$:
\beq\label{apksNP}
\Im\lambda_{\psi K_S}=\sin[2(\beta+\theta_d)].
\eeq
\item[(ii)] $\Im\lambda_{\phi K_S}$, the CP asymmetry in $B\rightarrow\phi K_S$.
This measurement will cleanly determine the relative phase between
the $B-\bar B$ mixing amplitude and the $b\ra s\bar ss$ decay amplitude.
The $b\ra s\bar ss$ decay has only Standard Model penguin contributions
and therefore is sensitive to new physics. We parametrize the modification
of the decay amplitude by a phase $\theta_A$ \cite{GrWo}:
\beq\label{aphksNP}
\Im\lambda_{\phi K_S}=\sin[2(\beta+\theta_d+\theta_A)].
\eeq
\item[(iii)] $a_{\pi\nu\bar\nu}$, the CP violating ratio of $K\rightarrow
\pi\nu\bar\nu$ decays:
\beq\label{defapnn}
a_{\pi\nu\bar\nu}={\Gamma(K_L\ra\pi^0\nu\bar\nu)\over
\Gamma(K^+\ra\pi^+\nu\bar\nu)}.
\eeq
This measurement will cleanly determine the relative phase between the 
$K-\bar K$ mixing amplitude and the $s\ra d\nu\bar\nu$ decay amplitude.
The experimentally measured small value of $\epsK$ requires that
the phase of the $K-\bar K$ mixing amplitude is not modified from
the Standard Model prediction. (More precisely, it requires that
the phase in the mixing amplitude is very close to the phase in
the $s\rightarrow d\bar uu$ decay amplitude.) On the other hand, the decay,
which in the Standard Model is a loop process with small mixing angles,
can be easily modified by new physics.
\item[(iv)] $\Im(\lambda_{K^-\pi^+})$, the CP violating
quantity in $D\rightarrow K^-\pi^+$ decay. The time-dependent
ratios between the doubly-Cabibbo-suppressed and the
Cabibbo-allowed $D\ra K\pi$ decay rates are given by
\beqa\label{deflD}
{\Gamma[D^0(t)\ra K^+\pi^-]\over
\Gamma[D^0(t)\ra K^-\pi^+]}\simeq&\ |\lambda|^2+
{(\Delta m_D)^2\over4}t^2+\Im(\lambda^{-1}_{K^+\pi^-})t,\\
{\Gamma[\bar D^0(t)\ra K^-\pi^+]\over
\Gamma[\bar D^0(t)\ra K^+\pi^-]}\simeq&\ |\lambda|^2+
{(\Delta m_D)^2\over4}t^2+\Im(\lambda_{K^-\pi^+})t,
\eeqa
where we used the fact that CP violation in mixing is expected
to be small and, consequently,
\beq\label{lKpi}
|\lambda^{-1}_{K^+\pi^-}|=|\lambda_{K^-\pi^+}|\equiv\lambda.
\eeq
The ratio
\beq\label{defaD}
a_{D\ra K\pi}={\Im(\lambda_{K^-\pi^+})\over|\lambda_{K^-\pi^+}|}
\eeq
depends on the relative phase between the $D-\bar D$ mixing amplitude
and the $c\ra d\bar su$ decay amplitude. Within the Standard Model,
this decay channel is tree level. It is unlikely that it is
affected by new physics. On the other hand, the mixing amplitude can
be easily modified by new physics \cite{BKS}.
\item[(v)] $d_N$, the electric dipole moment of the neutron.
We did not discuss this quantity so far because, unlike
CP violation in meson decays, flavor changing couplings
are not necessary for $d_N$. In other words, the CP violation
that induces $d_N$ is {\it flavor diagonal}. It does in general
get contributions from flavor changing physics, but it could
be induced by sectors that are flavor blind. Within the
Standard Model (and ignoring the strong CP angle $\theta_{\rm QCD}$),
the contribution from $\delta_{\rm KM}$ arises at the three loop
level and is at least six orders of magnitude below the
experimental bound \cite{PDG}\ $d_N^{\exp}$,
\beq\label{dnexp}
d_N^{\rm exp}=1.1\times10^{-25}\ e\ {\rm cm}.
\eeq
\end{itemize}

The various CP violating observables discussed above are sensitive
then to new physics in the mixing amplitudes for the $B-\bar B$
and $D-\bar D$ systems, in the decay amplitudes for $b\rightarrow s\bar ss$
and $s\ra d\nu\bar\nu$ channels and to flavor diagonal CP violation.
If information about all these processes becomes available
and deviations from the Standard Model predictions are found,
we can ask rather detailed questions about the nature of
the new physics that is responsible to these deviations:
\begin{itemize}
\item[(i)] Is the new physics related to the down sector?
the up sector? both?
\item[(ii)] Is the new physics related to $\Delta B=1$ processes?
$\Delta B=2$? both?
\item[(iii)] Is the new physics related to the third generation?
to all generations?
\item[(iv)] Are the new sources of CP violation flavor changing?
flavor diagonal? both?
\end{itemize}

It is no wonder then that with such rich information,
flavor and CP violation provide an excellent probe of new physics.

\section{Supersymmetry}
A generic supersymmetric extension of the Standard Model contains a host
of new flavor and CP violating parameters. (For reviews on supersymmetry see
refs. 
\cite{nilles,haberkane,barbieri,HaberTASI,RamoTASI,BaggTASI,SMart,LBH}. 
The following chapter is based on \cite{GNR}.) 
The requirement of consistency with experimental data provides strong 
constraints on many of these parameters. For this reason, the physics of flavor
and CP violation has had a profound impact on supersymmetric model building. 
A discussion of CP violation in this context can hardly avoid addressing the 
flavor problem itself.  Indeed, many of the supersymmetric models that we
analyze below were originally aimed at solving flavor problems.
 
As concerns CP violation, one can distinguish two classes of experimental
constraints. First, bounds on nuclear and atomic electric dipole moments
determine what is usually called the {\it supersymmetric CP problem}.
Second, the physics of neutral mesons and, most importantly, the small
experimental value of $\epsK$ pose the {\it supersymmetric $\epsK$
problem}. In the next two subsections we describe the two problems.
Then we describe various supersymmetric flavor problems and
the ways in which they address the supersymmetric CP problem.
 
Before turning to a detailed discussion, we define two scales that
play an important role in supersymmetry: $\Lambda_S$, where
the soft supersymmetry breaking terms are generated, and $\Lambda_F$,
where flavor dynamics takes place. When $\Lambda_F\gg\Lambda_S$, it is
possible that there are no genuinely new sources of flavor and CP
violation. This leads to models with exact universality, which we
discuss in section IV.C. When $\Lambda_F\lsim\Lambda_S$, we do not
expect, in general, that flavor and CP violation are limited to the
Yukawa matrices. One way to suppress CP violation would be to assume
that CP is an approximate symmetry of the full theory (namely, CP
violating phases are all small). We discuss this scenario in section IV.D.
Another option is to assume that, similarly to the Standard Model,
CP violating phases are large, but their effects are screened, possibly
by the same physics that explains the various flavor puzzles.
Such models, with Abelian or non-Abelian horizontal symmetries,
are described in section IV.E. It is also
possible that CP violating effects are suppressed because squarks
are heavy. This scenario is also discussed in section IV.E.
Some concluding comments are given in section IV.F.

\subsection{The Supersymmetric CP Problem}
 One aspect of supersymmetric CP violation involves effects that are
flavor preserving. Then, for simplicity, we describe this aspect in
a supersymmetric model without additional flavor mixings, {\it i.e.} the
minimal supersymmetric standard model (MSSM) with universal sfermion
masses and with the trilinear SUSY-breaking scalar couplings
proportional to the corresponding Yukawa couplings. (The generalization
to the case of non-universal soft terms is straightforward.)  In such a
constrained framework, there are four new phases beyond the two
phases of the Standard Model ($\delta_{\rm KM}$ and $\theta_{\rm QCD}$).
One arises in the bilinear $\mu$-term of the superpotential,
\beq\label{muterm}
W=\mu H_uH_d,
\eeq
while the other three arise in the soft supersymmetry breaking
parameters $m_\tgl$ (the gaugino mass), $A$ (the trilinear
scalar coupling) and $m_{12}^2$ (the bilinear scalar coupling):
\beq\label{sSUSYb}
\L=-{1\over2}m_\tgl\tgl\tgl-A(Y^u QH_u\bar u
-Y^d QH_d\bar d-Y^e LH_d\bar\ell)-m_{12}^2 H_uH_d+{\rm h.c.},
\eeq
where $\tgl$ are the gauginos and $Y$ are Yukawa matrices.
Only two combinations of the four phases are physical \cite{DGH,DiTh}.
In the absence of (\ref{muterm}) and (\ref{sSUSYb}), there are two additional 
global $U(1)$ symmetries in the MSSM, an $R$ symmetry and a Peccei-Quinn
symmetry. This means that one could treat the various dimensionful
parameters in (\ref{muterm}) and (\ref{sSUSYb}) as spurions which break the
symmetries, thus deriving selection rules. The appropriate
charge assignments are:
\beq\label{spur}
\matrix{&m_\tgl&A&m_{12}^2&\mu&H_u&H_d&Q\bar u&Q\bar d&L\bar\ell\cr
U(1)_{\rm PQ}&0&0&-2&-2&1&1&-1&-1&-1\cr
U(1)_{\rm R}&-2&-2&-2&0&1&1&1&1&1\cr}
\eeq
Physical observables can only depend on combinations of the dimensionful
parameters that are neutral under both $U(1)$'s. There are three such
independent combinations: $m_\tgl\mu(m_{12}^2)^*$, $A\mu(m_{12}^2)^*$
and $A^* m_\tgl$. However, only two of their phases are independent, say
\beq\label{phiAB}
\phi_A=\arg(A^* m_\tgl),\ \ \ \phi_B=\arg(m_\tgl\mu(m_{12}^2)^*).
\eeq
In the more general case of non-universal soft terms there is
one independent phase $\phi_{A_{i}}$ for each quark and lepton flavor.
Moreover, complex off-diagonal entries in the sfermion
mass matrices may represent additional sources of CP violation.
 
The most significant effect of $\phi_A$ and $\phi_B$ is their
contribution to electric dipole moments (EDMs). For example, the
contribution from one-loop gluino diagrams to the down quark EDM
is given by \cite{BuWy,PoWi}:
\beq\label{ddsusy}
d_d=M_d{e\alpha_3\over 18\pi\tilde m^4}\left(
|A m_{\tilde g}|\sin\phi_A+\tan\beta|\mu m_{\tilde g}|\sin\phi_B\right),
\eeq
where we have taken $m^2_Q\sim m^2_D\sim m^2_{\tilde g}\sim\tilde m^2$,
for left- and right-handed squark and gluino masses. We define, as usual,
$\tan\beta = \vev{H_u}/\vev{H_d}$. Similar one-loop diagrams give rise to
chromoelectric dipole moments. The electric and chromoelectric dipole moments 
of the light quarks $(u,d,s)$ are the main source of $d_N$ (the EDM of the 
neutron), giving \cite{FPT}\
\beq\label{dipole}
d_N\sim 2\, \left({100\, \gev\over \tilde m}\right )^2
\sin \phi_{A,B}\times10^{-23}\ e\, {\rm cm}
\eeq
where, as above, $\tilde m$ represents the overall SUSY scale.
The present experimental bound, $d_N<1.1\times 10^{-25}e\, {\rm cm}$
\cite{smith,altarev}, is then violated for  $\O(1)$ phases, unless the
masses of superpartners are above $\O(1\ TeV)$. Alternatively for light
SUSY masses, the new phases should be $<\O(10^{-2})$. Notice however that
one may consider the actual bound weaker than this, due to the
theoretical uncertainty in the estimate of the hadronic matrix elements
that lead to eq. (\ref{dipole}) \cite{florellis}. With this caveat, whether
the phases are small or squarks are heavy, a fine-tuning of order
$10^{-2}$ seems to be required, in general, to avoid too large a $d_N$.
This is {\it the Supersymmetric CP Problem} \cite{BuWy,PoWi,EFN,Barr}.
 
In addition to $d_N$, the SUSY CP phases contribute to atomic and nuclear
EDMs (see a detailed discussion in ref. \cite{FPT}).
The former are also sensitive to phases in the leptonic sector.
The latter give additional constraints on the quark sector phases.
For instance, the bound on the nuclear EDM of ${}^{199} {\rm Hg}$
is comparable to the one given by $d_N$.
 In practice, these additional bounds on SUSY CP phases
are not stronger than those from $d_N$ (at least in the quark sector).
However, since there are significant theoretical uncertainties in the
calculation of nuclear EDMs, it is important to measure as many as
possible of them to obtain more reliable bounds. 

\subsection{The Supersymmetric $\epsK$ Problem}
 The contribution to the CP violating $\epsK$ parameter in the neutral $K$
system is dominated by diagrams involving $Q$ and $\bar d$ squarks in the
same loop \cite{DNW,GaMa,HKT,GMS,GGMS}. The corresponding effective four-fermi 
operator involves fermions of both chiralities, so that its matrix elements
are enhanced by $\O(m_K/m_s)^2$ compared to the chirality conserving
operators. For $m_{\tilde g}\simeq m_Q \simeq m_D= \tilde m$ (our results
depend only weakly on this assumption) and focusing on the contribution
from the first two squark families, one gets (we use the results in
ref. \cite{GGMS})
\beq\label{epsKSusy}
\epsK={5\ \alpha_3^2  \over 162\sqrt2}{f_K^2m_K\over\tilde
m^2\Delta m_K}\left [\left({m_K\over m_s+m_d}\right)^2+{3\over 25}\right]
\Im\left\{{(\delta m_Q^2)_{12}\over m_Q^2}
{(\delta m_D^2)_{12}\over m_D^2}\right\},
\eeq
where $(\delta m_{Q,D}^2)_{12}$ are the off diagonal entries in the
squark mass matrices in a basis where the down quark mass matrix
and the gluino couplings are diagonal. These flavor
violating quantities are often written as
$(\delta m_{Q,D}^2)_{12}=V_{11}^{Q,D}\delta m_{Q,D}^2 V_{21}^{Q,D*}$,
where $\delta m_{Q,D}^2$ is the mass splitting among the squarks
and $V^{Q,D}$ are the gluino coupling mixing matrices in the mass
eigenbasis of quarks and squarks. Note that CP would be violated
even if there were two families only \cite{NirSusy}. (There are also 
contributions involving the third family squarks via the (13) and (23) 
mixings. In some cases the third family contribution actually dominates.) 
Using the experimental value of $\epsK$, we get
\beq\label{epsKScon}
{(\Delta m_K\epsK)^{\rm SUSY}\over(\Delta m_K\epsK)^{\rm EXP}}\sim10^7
\left ({300 \ \gev\over\tilde m}\right)^2
\left({m^2_{Q_2}-m^2_{Q_1}\over m_Q^2}\right)
\left({m^2_{D_2}-m^2_{D_1}\over m_D^2}\right)
|K_{12}^{dL}K_{12}^{dR}|\sin\phi,
\eeq
where $\phi$ is the CP violating phase. In a generic supersymmetric framework, 
we expect $\tilde m=\O(m_Z)$, $\delta m_{Q,D}^2/m_{Q,D}^2=\O(1)$, 
$K_{ij}^{Q,D}=\O(1)$ and $\sin\phi=\O(1)$. Then the constraint 
(\ref{epsKScon}) is generically violated by about seven orders of magnitude. 
(Four-fermi operators with same chirality fermions give a smaller effect. 
The resulting $\epsK$-bounds are therefore weaker by about one order of 
magnitude.)

Eq. (\ref{epsKScon}) also shows what are the possible ways to solve
the supersymmetric $\epsK$ problem:
\begin{itemize}
\item[(i)] Heavy squarks: $\tilde m\gg300\ GeV$;
\item[(ii)] Universality: $\delta m_{Q,D}^2\ll m_{Q,D}^2$;
\item[(iii)] Alignment: $|K_{12}^d|\ll1$;
\item[(iv)] Approximate CP: $\sin\phi\ll1$.
\end{itemize}
The $d_N$ problem (see eq. (\ref{dipole}))
is solved by either heavy squarks or approximate CP.

\subsection{Exact Universality}
Both supersymmetric CP problems are solved if, at the scale $\Lambda_S$,
the soft supersymmetry breaking terms are universal and the genuine SUSY
CP phases $\phi_{A,B}$ vanish. Then the Yukawa matrices represent
the only source of flavor and CP violation which is relevant in low
energy physics. This situation can naturally arise when supersymmetry
breaking is mediated by gauge interactions at a scale $\Lambda_S\ll\Lambda_F$ 
\cite{DiNe,DNeS,DNNS,DNiS}. In the simplest scenarios, the $A$-terms and the
gaugino masses are generated by the same SUSY and $U(1)_R$ breaking
source (see eq. (\ref{spur}). Thus, up to very small effects due to the
{\it standard}  Yukawa matrices, $\arg(A)=\arg(m_{\tilde g})$ so that
$\phi_A$ vanishes. In specific models also $\phi_B$ vanishes in a similar
way \cite{DNeS,DNiS}. It is also possible that similar boundary
conditions occur when supersymmetry breaking is communicated to the
observable sector up at the Planck scale \cite{CAN,BFS,HLW,EKN,LaRo}. 
The situation in this case seems to be less under control from the theoretical 
point of view. Dilaton dominance in SUSY breaking, though, seems a very 
interesting direction to explore \cite{KaLo,BML}.
 
The most important implication of this type of boundary conditions
for soft terms, which we refer to as {\it exact  universality}
\cite{DiGe,Saka}, is the existence of the SUSY analogue of the GIM mechanism 
which operates in the SM. The CP violating phase of the CKM matrix can
feed into the soft terms via Renormalization Group (RG) evolution only
with a strong suppression from light quark masses \cite{DGH}.
 
With regard to the supersymmetric CP problem, gluino diagrams
contribute to quark EDMs as in eq. (\ref{ddsusy}),
but with a highly suppressed effective phase, {\it e.g.}
\beq\label{gim}
\phi_{A_d}\sim (t_S/16 \pi^2)^4 Y_t^4 Y_c^2 Y_b^2 J.
\eeq
Here $t_S=\log (\Lambda_S/M_W)$ arises from the RG evolution from
$\Lambda_S$ to the electroweak scale, the $Y_i$'s are quark Yukawa
couplings (in the mass basis), and $J\simeq 2\times 10^{-5}$ is defined in eq. 
(\ref{defJ}). A similar contribution comes from chargino diagrams. The 
resulting EDM is $d_N\lsim 10^{-31}\ e\ {\rm cm}$. This maximum can be reached 
only for very large $\tan\beta\sim60$ while, for small $\tan\beta \sim 1$,
$d_N$ is about 5 orders of magnitude smaller. This range of values for
$d_N$ is much below the present ($\sim10^{-25}\ e$ cm) and foreseen
($\sim 10^{-28}\ e$ cm) experimental sensitivities \cite{RoStru,BeVi,IMSS,ACW}.
 
With regard to the supersymmetric $\epsK$ problem, the contribution
to $\epsK$ is proportional to $\Im(V_{td}V_{ts}^*)^2 Y_t^4
(t_S/16 \pi^2)^2$, giving the same GIM suppression as in the SM.
This contribution turns out to be small \cite{DGH}:
\beq\label{unieps}
|\epsK^{\rm SUSY}|\sim 6 \times 10^{-6} \left [{J\,\Re
(V_{td}V_{ts}^*)\over10^{-8}}\right]\left[{300\gev\over\tilde m}\right]^2
\left [ {\ln (\Lambda_S/m_W)\over 5}\right ]^2.
\eeq
The value $t_S=5$ is typical to gauge mediated supersymmetry breaking,
but (\ref{unieps}) remains negligible for any scale $\Lambda_S\lsim
M_{\rm Pl}$ (namely $t_S\lsim 35$). The supersymmetric contribution to
$D-\bar D$ mixing is similarly small and we expect no observable effects.
 
For the $B_d$ and $B_s$ systems, the largest SUSY contribution
to the mixing comes from box diagrams with intermediate charged Higgs and
the up quarks. It can be up to $\O(0.2)$ of the SM amplitude for
$\Lambda_S=M_{\rm Pl}$ and $\tan\beta = \O(1)$ \cite{BBMR,GNO,Nihe,BaKo},
and much smaller for large $\tan\beta$. The contribution is smaller in
models of gauge mediated SUSY breaking where the mass of the charged
Higgs boson is typically $\gsim 300\ GeV$ \cite{DNNS}\ and $t_S\sim5$. The
SUSY contributions to $B_s-\bar B_s$ and $B_d -\bar B_d$ mixing are, to a
good approximation, proportional to $(V_{tb}V_{ts}^*)^2$ and
$(V_{tb}V_{td}^*)^2$, respectively, like in the SM. Then, regardless of
the size of these contributions, the relation $\Delta m_{B_d}/\Delta m_{B_s}
\sim |V_{td}/V_{ts}|^2$ and the CP asymmetries in neutral
$B$ decays into final CP eigenstates are the same as in the SM.

\subsection{Approximate CP Symmetry}
Both supersymmetric CP problems are solved if CP is an approximate
symmetry, broken by a small parameter of order $10^{-3}$. This is one
of the possible solutions to CP problems in the class of supersymmetric
models with $\Lambda_F\lsim\Lambda_S$, where the soft masses are
generically not universal, so that we do not expect flavor and CP
violation to be limited to the Yukawa matrices. (Of course,
some mechanism has also to suppress the real part of the $\Delta S=2$
amplitude by a sufficient amount.) Most models where soft terms arise at
the Planck scale ($\Lambda_S\sim M_{\rm Pl}$) belong to this class.
 
If CP is an approximate symmetry, we expect also the SM
phase $\delta_{\rm KM}$ to be $\ll 1$. Then the standard box diagrams
cannot account for $\epsK$ which should arise from another
source. In supersymmetry with non-universal soft terms, the source could
be diagrams involving virtual superpartners, mainly squark-gluino box
diagrams. Let us call $(M_{12}^K)^{\rm SUSY}$
the supersymmetric contribution to the $K-\bar K$ mixing amplitude.
Then the requirements $\Re (M_{12}^K)^{\rm SUSY}\lsim\Delta m_K$
and $\Im(M_{12}^K)^{\rm SUSY}\sim\epsK\Delta m_K$ imply that the
generic CP phases are $\geq\O(\epsK)\sim 10^{-3}$.

Of course, $d_N$ constrains the relevant CP violating phases to be
$\lsim10^{-2}$. If all phases are of the same order, then $d_N$ must be
just below or barely compatible with the present experimental bound.
A signal should definitely be found if the accuracy is increased by two
orders of magnitude.
 
The main phenomenological implication of these scenarios is that
CP asymmetries in $B$ meson decays are small, perhaps $\O(\epsK)$, rather 
than ${\cal O}(1)$ as expected in the SM. Also the ratio $\apnn$ (see 
(\ref{defapnn})) is very small, in contrast to the Standard Model
where it is expected to be of $O(\sin^2\beta)$. Explicit models of 
approximate CP were presented in refs. \cite{abefre,BaBa,Eyal}.  

The fact that the Standard Model and the models of approximate
CP are both viable at present is related to the fact that the
mechanism of CP violation has not really been tested experimentally.
The only measured CP violating observale, that is $\epsK$, is small.
Its smallness could be related to the `accidental' smallness of CP violation
for the first two quark generations, as is the case in the Standard
Model, or to CP being an approximate symmetry, as is the case in the
models discussed here. Future measurements, particularly of processes
where the third generation plays a dominant role (such as $\apks$
or $\apnn$), will easily distinguish between the two scenarios.
While the Standard Model predicts large CP violating effects
for these processes, approximate CP would suppress them too.
 
The distinction between the Standard Model and Supersymmetry could
also be made -- though less easily -- in measurements of CP violation
in neutral $D$ decays and of the electric dipole moments of the
neutron. Here, the GIM mechanism of the Standard Model is so
efficient that CP violating effects are unobservable in both cases.
In contrast, the flavor breaking in supersymmetry might be much
stronger, and then the approximate CP somewhat suppresses the effects
but to a level which is perhaps still observable.

\subsection{Approximate Horizontal Symmetries}
Another option is to assume that, similarly to the Standard Model,
CP violating phases are large, but their effects are screened, possibly
by the same physics that explains the various flavor puzzles. This
usually requires Abelian or non-Abelian horizontal symmetries.
Two ingredients play a major role here: selection rules that come from
the symmetry and holomorphy of Yukawa and $A$-terms that comes from the
supersymmetry. With Abelian symmetries, the screening mechanism is
provided by {\it alignment} \cite{NiSe,LNSb,RaNi}, whereby the mixing matrices 
for gaugino couplings have very small mixing angles, particularly for the
first two down squark generations. With non-Abelian symmetries, the
screening mechanism is {\it approximate universality}, where squarks of
the two families fit into an irreducible doublet and are,
therefore, approximately degenerate 
\cite{DKL,PoSe,HaMu,PoTo,BDH,CHM,Zurab,Raby,Gali}. In all of these
models, it is difficult to avoid $d_N\gsim10^{-28}$ e cm.
 
As far as the third generation is concerned, the signatures of
Abelian and non-Abelian models are similar. In particular, they allow
observable deviations from the SM predictions for CP asymmetries in
$B$ decays. In some cases, non-Abelian models give relations between
CKM parameters and consequently predict strong constraints
on these CP asymmetries.

For the two light generations, only alignment allows
interesting effects. In particular, it predicts large CP violating
effects in $D-\bar D$ mixing \cite{NiSe,LNSb}. Thus, it allows for
$a_{D\rightarrow K\pi}=\O(1)$ and, in particular, for
$\Im(\lambda^{-1}_{K^+\pi^-})\neq \Im(\lambda_{K^-\pi^+})$
which will signify new CP violating phases (see eqs. (\ref{deflD})).
 
Finally, it is possible that CP violating effects are suppressed because
squarks are heavy. If the masses of the first and second generations
squarks $m_i$ are larger than the other soft masses, $m_i^2\sim 100\,
\tilde m^2$ then the Supersymmetric CP problem is solved and the $\epsK$
problem is relaxed (but not eliminated) \cite{PoTo,DKL}. This does not
necessarily lead to naturalness problems, since these two generations are
almost decoupled from the Higgs sector.

Notice though that, with the possible exception of $m_{\tilde b_R}^2$,
third family squark masses cannot naturally be much above $m_Z^2$.
If the relevant phases are of $O(1)$, the main contribution to $d_N$
comes from the third family via the two-loop induced three-gluon operator
\cite{Weintg}, and it is roughly at the present experimental
bound  when $m_{\tilde t_{L,R}}\sim 100\ GeV$.
 
Models with the first two squark generations heavy have their own
signatures of CP violation in neutral meson mixing \cite{CKLN}.
The mixing angles relevant to $D-\bar D$ mixing are similar, in general,
to those of models of alignment (if alignment is
invoked to explain $\Delta m_K$ with $m^2_{Q,D}\lsim20\ TeV$).
However, since the $\tilde u$ and $\tilde c$ squarks are heavy, the
contribution to $D-\bar D$ mixing is one to two orders
of magnitude below the experimental bound. This may lead to the
interesting situation that $D-\bar D$ mixing will first be observed
through its CP violating part \cite{WolfD}.
In the neutral $B$ system, $\O(1)$ shifts from the Standard Model
predictions of CP asymmetries in the decays to final CP eigenstates
are possible. This can occur even when the squarks masses of the third
family are $\sim1\ TeV$ \cite{CKNS}, 
since now mixing angles can naturally be larger than in the case 
of horizontal symmetries (alignment or approximate universality).

\subsection{Some Concluding Comments}
 
The conclusion from our discussion of supersymmetric flavor models is
that measurements of CP violation will provide us with an excellent probe 
of the flavor and CP structure of supersymmetry. This is clearly demonstrated 
in Table I. 
\begin{table}
\[ \begin{array}{|c|c|c|c|c|c|}
\hline
{\rm Model} & d_N/d_N^{\rm exp} & \theta_d & \theta_A &
a_{D^0\ra K^-\pi^+} & a_{K\ra\pi\nu\bar\nu} \\
\hline
{\rm Standard\ Model} & \lsim10^{-6} & 0 & 0 & 0 & \O(1) \\
{\rm Exact Universality} & \lsim10^{-6} & 0 & 0 & 0 & ={\rm SM}  \\
{\rm Approximate\ CP} & \sim10^{-1} & -\beta & 0 & \O(10^{-3}) 
& \O(10^{-5}) \\
{\rm Alignment} & \gsim10^{-3} & \O(0.2) & \O(1) &
\O(1) & \approx{\rm SM} \\
{\rm Approx.\ Universality} & \gsim10^{-2} & \O(0.2) & \O(1) &
0 & \approx{\rm SM} \\
{\rm Heavy\ Squarks} & \sim10^{-1} & \O(1) & \O(1) &
\O(10^{-2}) & \approx{\rm SM} \\
\hline
\end{array} \]
\vskip 12pt
\caption
{CP violating observables in various classes
of Supersymmetric flavor models.}
\end{table}

The unique features of CP violation are well demonstrated by
examining the CP asymmetry in $B\rightarrow\psi K_S$,
$\Im\lambda_{\psi K_S}$, and CP violation in $K\rightarrow\pi
\nu\bar\nu$, $\Im\lambda_{\pi\nu\bar\nu}$. Model independently,
$\Im\lambda_{\psi K_S}$ measures the relative phase between the
$B-\bar B$ mixing amplitude and the $b\rightarrow c\bar cd$ decay
amplitude (more precisely, the $b\rightarrow c\bar cs$ decay
amplitude times the $K-\bar K$ mixing amplitude), while
$\Im\lambda_{\pi\nu\bar\nu}$ measures the relative phase between the
$K-\bar K$ mixing amplitude and the $s\rightarrow d\nu\bar\nu$ decay
amplitude. We would like to emphasize the following three points:

\begin{itemize}
\item[(i)] {\it The two measurements are theoretically clean to better
than} $\O(10^{-2})$. Thus they can provide the most accurate
determination of CKM parameters.
\item[(ii)] {\it As concerns CP violation, the Standard Model is a uniquely
predictive model}. In particular, it predicts that the seemingly
unrelated $\Im\lambda_{\psi K_S}$ and $\Im\lambda_{\pi\nu\bar\nu}$
measure the same parameter, that is the angle $\beta$ of the
unitarity triangle.
\item[(iii)] {\it In the presence of New Physics, there is in general no
reason for a relation between $\Im\lambda_{\psi K_S}$ and
$\Im\lambda_{\pi\nu\bar\nu}$}. Therefore, a measurement of both
will provide a sensitive probe of New Physics.
\end{itemize}

\bigskip
\noindent
{\bf Acknowledgements:}
\smallskip
\noindent
My understanding of CP violation has benefitted
from numerous discussions with Helen Quinn. Large parts of this
review are based on previous reviews written in collaboration
with her. Parts of this review are based on contributions
to the BaBar Physics Book that were written in collaboration
with Gerald Eigen, Ben Grinstein, Stephane Plaszczynski and
Marie-Helene Schune. Other parts of this review are based
on a review written in collaboration with Yuval Grossman and
Riccardo Rattazzi. Special thanks go to Stephane Plaszczynski and
Marie-Helene Schune for performing the CKM fit and preparing the plots for
this review and to Joao Silva for explaining to me various subtleties
concerning phase conventions and for useful comments on the manuscript. 
This manuscript is based on sets
of lectures given in the school on flavor and gauge hierarchies
(Cargese, 1998) and in the school on B and CP within and beyond
the Standard Model (KIAS, Seoul, 1999). I thank Pierre Binetruy,
and Seongyoul Choi and Kiwoon Choi for their hospitality in
the respective schools. 
Y.N. is supported in part by the United States $-$ Israel Binational
Science Foundation (BSF) and by the Minerva Foundation (Munich).

\end{document}